# Discovering novel systemic biomarkers in photos of the external eye

**Authors**:
Boris Babenko PhD[1†], Ilana Traynis MD[2], Christina Chen MD[1], Preeti Singh MS[1], Akib Uddin MHSc[1], Jorge Cuadros OD PhD[3], Lauren P. Daskivich MD MSHS[4,5], April Y. Maa MD[6,7], Ramasamy Kim MD[8], Eugene Yu-Chuan Kang MD[9], Yossi Matias PhD[1], Greg S. Corrado PhD[1], Lily Peng MD PhD[1], Dale R. Webster PhD[1], Christopher Semturs MS[1], Jonathan Krause PhD[1], Avinash V. Varadarajan MS[1], Naama Hammel MD[1†]*, Yun Liu PhD[1†]*

[1]Google Health, 3400 Hillview Ave. Palo Alto, CA, USA
[2]Work done at Google Health via Advanced Clinical, 6 Parkway N Ste 100, Deerfield, IL, USA
[3]EyePACS Inc, 211 Chace St, Santa Cruz, CA, USA
[4]Ophthalmic Services and Eye Health Programs, Los Angeles County Department of Health Services, 313 N. Figueroa St #904C, Los Angeles, CA, USA
[5]Department of Ophthalmology, University of Southern California Keck School of Medicine/Roski Eye Institute, 1450 San Pablo Street, Los Angeles, CA USA
[6]Department of Ophthalmology, Emory University School of Medicine, 1365 Clifton Rd B, Atlanta, GA USA
[7]Regional Telehealth Services, Technology-based Eye Care Services (TECS) division, Veterans Integrated Service Network (VISN) 7, Decatur, GA, USA
[8]Aravind Eye Hospital, Aravind, Eye Hospital, N0.1, Anna Nagar, Madurai-625020, Tamil Nadu, India
[9]Department of Ophthalmology, Linkou Medical Center, Chang Gung Memorial Hospital, Taoyuan, Taiwan

[†]These authors contributed equally
*Corresponding authors, nhammel@google.com, liuyun@google.com

# Abstract

**Background**
Photographs of the external eye were recently shown to reveal signs of diabetic retinal disease and elevated HbA1c. In this paper, we evaluate if external eye photos contain information about additional systemic medical conditions.

**Methods**
We developed a deep learning system (DLS) that takes external eye photos as input and predicts multiple systemic parameters, such as those related to the liver (albumin, aspartate aminotransferase [AST]); kidney (glomerular filtration rate estimated using the race-free 2021 CKD-EPI creatinine equation [eGFR], the urine albumin/creatinine ratio [ACR]); bone & mineral (calcium); thyroid (thyroid stimulating hormone, TSH); and blood count (hemoglobin [Hgb], white blood cells [WBC], platelets). This network was developed using 151,237 images from 49,015 patients with diabetes undergoing diabetic eye screening in 11 sites across Los Angeles county, CA. Evaluation focused on 9 pre-specified systemic parameters and leveraged 3 validation sets (A, B, C) spanning 28,869 patients with and without diabetes undergoing eye screening in 3 independent sites in Los Angeles county, CA, and the greater Atlanta area, GA. To ensure the DLS did not solely depend on demographic information extractable from external eye photos (or easily obtainable from asking the patient), we compared against “baseline” models incorporating available clinicodemographic variables (e.g. age, sex, race/ethnicity, years with diabetes).

**Results**
Relative to the baseline, the DLS achieved statistically significant superior performance at detecting AST > 36.0 U/L, calcium < 8.6 mg/dL, eGFR < 60.0 mL/min/1.73 $m^2$, Hgb < 11.0 g/dL, platelets < 150.0 $10^3$/μL, ACR ≥ 300.0 mg/g, and WBC < 4.0 $10^3$/μL on validation set A (representing a patient population similar to the development datasets), where the AUC of DLS exceeded that of the baseline model by 5.2 to 19.4%. On validation sets B and C, with substantial patient population differences compared to the development datasets (e.g. including patients with and without diabetes), the DLS outperformed the baseline for ACR ≥ 300.0 mg/g and Hgb < 11.0 g/dL by 7.3 to 13.2%.

**Conclusion**
Our findings provide further evidence that external eye photos contain important biomarkers of systemic health spanning multiple organ systems. Further work is needed to investigate whether and how these biomarkers can be translated into clinical impact.

# Introduction

Ocular sequelae resulting from systemic disease have been well documented[1,2] and are the basis for globally established screening programs that identify diabetic retinal disease from retinal fundus photography[3]. More recent work has shown that a number of systemic biomarkers, such as blood pressure, glycosylated hemoglobin levels[4], and estimated glomerular filtration rate[5], can be detected from fundus photographs as well. Although this suggests an opportunity for non-invasive detection of systemic disease, the need for a trained photographer and a specialized fundus camera to capture retinal photos presents a barrier to clinical implementation.

By contrast, our previous work[6] examined whether a few select markers of systemic and ocular disease related to diabetes could be identified from photos of the external eye, which in principle does not require specialized cameras to obtain. We found that our machine learning model could in fact identify poor blood glucose control, diabetic retinopathy, and diabetic macular edema from images of the external eye. Since clinical literature has established that external eye structures manifest many signs of systemic disease beyond diabetes, we considered whether machine learning could glean similar and perhaps novel signals of other systemic disease from photographs of the external eye.

In our current work we hypothesized that machine learning could predict several additional biomarkers of systemic diseases, such as ones related to kidney, liver, thyroid, blood count, and cardiovascular health, from external eye photos. To test this hypothesis we utilized data where patients had both external eye photos taken and clinical parameters or laboratory measurements (e.g. serum chemistry panels or complete blood counts). We investigated measurements spanning a variety of systems: albumin, albumin-to-creatinine ratio (ACR), calcium, estimated glomerular filtration rate (eGFR), aspartate transaminase (AST), thyroid stimulating hormone (TSH), hemoglobin (Hgb), platelets, white blood cell (WBC), blood pressure (BP), body mass index (BMI), and more. We found that abnormalities in a number of parameters (ACR, calcium, eGFR, AST, Hgb, platelets, and WBC) could be detected from external eye photos.

# Methods

**Datasets**

To develop and evaluate the deep learning system (DLS), we used data from three sources in this work: (1) clinics in the Los Angeles County Department of Health Services (LACDHS) utilizing the EyePACS teleretinal diabetic screening program (61,230 patients), (2) primary care clinics in the greater Atlanta area participating in the Technology-based Eye Care Services (TECS) program (10,402 patients), and (3) community-based outpatient clinics in the Atlanta VA Healthcare system's diabetic retinopathy screening program (6,266 patients; see Table 1).

From the EyePACS/LACDHS data, 11 sites were used for development, 4 for tuning, and 3 for external validation (validation dataset "A"). Patients who had visits at sites across these splits were excluded.

The two validation sets from the Atlanta area ("B" and "C") served as additional external validation. Most of the patients across all datasets had their pupils dilated in preparation for fundus photography (i.e. not dilated specifically for external eye photography). Ethics review and Institutional Review Board exemption for this retrospective study on de-identified data were obtained via Advarra, County of Los Angeles Public Health, and Emory University Institutional Review Boards.

At all sites, external eye photographs were taken as part of the standard imaging protocol for teleretinal eye screening to assess the anterior segment. The imaging protocols were largely similar at the LACDHS and VA sites, with the eyes imaged using the Topcon (NW8 and NW400) and Zeiss Cirrus Photo 600 fundus cameras, and the patient positioned slightly further from the camera relative to the position used for photographing the retinal fundus. . More details can be found in Appendix 10. None of the external eye images were excluded for image quality reasons.

**Baseline characteristics and parameters predicted**

The parameters predicted by the DLS included baseline characteristics and those requiring clinical or laboratory measurements. Baseline clinicodemographic characteristics (i.e. age, sex, race/ethnicity, years with diabetes) were self-reported by patients and recorded by each site. For systolic and diastolic blood pressure we averaged all measurements within 90 days, and the same was done for weight. Height was averaged over 365 days (Appendix 5 discusses the sensitivity of results with respect to these time spans).

Diabetes status for the VA TECS dataset was determined from both the International Classification of Diseases codes (ICD-10 codes E09-E11 and E13; ICD-9 codes 250 and 362.0) available in the medical records and notes from eye care visits (e.g. if the patient self-reported being diagnosed with diabetic retinopathy [DR], they were labeled as diabetic because the DR diagnosis requires a diabetes diagnosis). Intraocular lens status in the VA TECS dataset was determined from the notes of the eye care visit. Cataract presence in the EyePACS/LACDHS images was indicated by the LACDHS graders; this was not available in the VA datasets.

Serum and urine laboratory measurements, as well as vitals, were extracted from the patients' medical records. For serum and urine labs, we took the measurement closest in time to the acquisition of each photo, and excluded lab results that were more than 30 days away for international normalized ratio (INR), more than 90 days away for HbA1c, and more than 180 days away for all other measurements.

In all datasets we computed the eGFR using the race-free 2021 CKD-EPI Creatinine equation[7], as recommended by the National Kidney Foundation and the American Society of Nephrology's Task Force on Reassessing the Inclusion of Race in Diagnosing Kidney Disease, in the absence of cystatin C in retrospective data[8]. This also ensured consistency of eGFR estimation across datasets and over time.

Some labels (diabetic retinal diseases - DR, diabetic macular edema, and vision threatening DR - and smoking status) were used only as auxiliary training signals during model training. Diabetic retinal disease labels were collected in the same manner as reported in prior work[6]. Smoking status was self-reported by the patient.

**Deep learning system development**

We trained a convolutional neural network to take an external eye photograph as input, and predict all labs and vitals in a multi-task fashion (i.e. one prediction "head" per task). Specifically, each lab and vital was thresholded (cut-offs are listed in Supp. Table 5) to facilitate uniform treatment as classification tasks and use of the standard cross entropy loss during training. The cut-offs were selected during model development by consulting the American Board of Internal Medicine (ABIM) lab reference ranges[9] and taking into account dataset statistics (i.e. we eschewed cutoffs that resulted in too few positive cases to train and evaluate on reliably). Labs and vitals with multiple cutoffs were configured as a single multi-class head (e.g. a lab with two cutoffs [X and Y] was configured to have 3 classes, corresponding to the 3 categories: ≤X, >X and ≤Y, >Y).

From the list of all prediction targets, we selected a subset of the 9 most promising ones based on performance of the model versus the baseline model on the tune set, potential clinical utility, and representation of multiple physiological systems: albumin < 3.5 g/dL (liver/kidney), AST > 36.0 U/L (liver), calcium < 8.6 mg/dL (kidney/endocrine), eGFR < 60.0 mL/min/1.73 $m^2$ (kidney), Hgb < 11.0 g/dL (blood count), platelet < 150.0 $10^3$/μL (blood count), TSH > 4.0 mU/L (thyroid), urine albumin/creatinine ratio (ACR) ≥ 300.0 mg/g (kidney), WBC < 4.0 $10^3$/μL (blood count). We pre-specified these primary analyses and associated multiple testing corrections (see the Statistical analysis section) prior to running the model on any of the validation sets.

We used an ensemble of the best five models from a hyperparameter search as the final model (hyperparameters for these models are reported in Supp. Table 7). Ensembling and additional modeling details are discussed in the Appendix 1. To generate predictions for all test sets we averaged the predicted likelihoods across the models in the ensemble, and across the two eyes of a patient in a given visit, when available.

Additional supplementary analysis included a deep learning model that takes both an external eye and baseline characteristics as input (Appendix 2). For this model, we one-hot-encoded categorical metadata variables and normalized scalar values to unit variance and zero mean using the training set statistics, and then concatenated these to the prelogits (i.e. penultimate layer of the convolutional neural network).

**Deep learning system evaluation**

To evaluate our approach we computed the difference between the area under receiver operator characteristic curve (AUC) of the DLS and that of a baseline model that only takes baseline characteristics as input (detailed below). To avoid bias towards patients with many visits, for each prediction target, we first selected the visits where a measurement for the target was available, and then selected one random visit per patient (if there was more than one).

**Baseline models for comparison**
Baseline models that the DLS was compared against were logistic regression models trained with the scikit-learn Python library, using class-balanced weighting and the default L2 regularization setting (C=1.0). Each validation set had different variables available, with EyePACS/LACDHS being most comprehensive. To avoid having to drop a large fraction of data in our primary analysis, we used different baselines for each dataset, such that each baseline model used only input variables that were available for at least 85% of the data points in the respective dataset: age, sex, race/ethnicity and years with diabetes for EyePACS/LACDHS test; age, sex, and race/ethnicity for VA TECS; age and sex for VA TRI. All baseline models were trained on the EyePACS/LACDHS training dataset (same as the DLS). Secondary analysis includes comparison to baselines with additional variables (e.g. body mass index and blood pressure; see Appendix 8).

**Statistical analysis**
We measured the AUC and corresponding 95% confidence intervals for both the DLS and the baselines using the DeLong method[10]. To compute the statistical significance of superiority of the DLS over the baselines we used the DeLong paired AUC comparison test[10]. The superiority analyses on the 9 prediction tasks listed in the "Deep learning system development" section were prespecified and documented as primary analyses before analyzing performance on the validation sets. Alpha was adjusted using Bonferroni correction for multiple hypotheses testing ($\alpha$ = 0.05 for one-sided superiority test, divided by 9 tasks = 0.0056).

Prespecified secondary analyses included AUC comparison for the complete set of labs, vitals and corresponding cutoffs, subgroup analysis, additional metrics (e.g. PPV), explainability analysis and sensitivity analysis with respect to the time gap between lab or vital and the photo.

**Role of the funding source**
Google LLC was involved in the design and conduct of the study; management, analysis, and interpretation of the data; preparation, review, or approval of the manuscript; and decision to submit the manuscript for publication.

# Results

The three data sources spanned a variety of patient populations and imaging protocols (Table 1). The EyePACS/LACDHS dataset (including validation set A) and VA teleretinal screening dataset (validation set C) were exclusively patients with diabetes presenting for diabetic retinopathy screening, while the TECS dataset (validation set B) included patients both with and without diabetes. With respect to demographic variables, patients in validation set A had a mean age of about 57 years, slightly skewed towards female (55%), and the majority were Hispanic (80%). Validation sets B and C represent a veteran population, both with a mean age above 60 years and predominantly male (85-95%). In validation set C (where this information was available), patients were majority Black (53%) or White (41%). Below, results presented are for the external validation sets (A, B, and C).

**Detection performance for abnormal labs**

For the set of primary analyses, we first compared the AUC of the external eye model against the AUC of a baseline that takes baseline variables (without the external eye image) as input. Figure 1 summarizes these results for all three validation sets (see Supp. Table 1 for more complete data including p-values, and Supp. Figure 2 for full ROC curves for validation set A). We found that for ACR ≥ 300.0 mg/g (severely increased albuminuria) and Hgb < 11.0 g/dL (moderate anemia), the external eye model achieved AUCs between 66% and 82% depending on the dataset. These AUCs were consistently statistically significantly higher than the baseline models across all validation sets: absolute AUC improvements of 8.2%, 13.2% and 11.7% for ACR ≥ 300.0 mg/g in the three validation sets respectively ($p < 0.0001$ for all), and AUC improvements of 19.4% and 7.3% for Hgb < 11.0 g/dL in validation sets A and B ($p < 0.005$ for both; this label was not available in validation set C).

For the eGFR < 60.0 mL/min/1.73 $m^2$ task (equivalent to Stage 3 or worse chronic kidney disease), we found that the external eye model achieved AUCs of 72-81%. The AUC was significantly higher than the baseline (by 8.5%, $p < 0.0001$) in validation set A, but not in either validation sets B or C. In our secondary analysis, we found that in the more severe cutoff eGFR < 30 mL/min/1.73 $m^2$ task (equivalent to Stage 4 or worse chronic kidney disease), the external eye model achieved AUCs of 77-88%, which outperformed the baseline across all three validation sets (AUC improvements from 10.7-17.0%, $p < 0.0005$ for all).

For albumin < 3.5 g/dL, AST > 36.0 U/L, and calcium < 8.6 mg/dL tasks, the external eye model achieved AUCs of 62-77%, significantly higher than the baseline (by 12.9%, 5.2%, and 9.2%, respectively, $p < 0.0001$ for all) in validation set A. These variables were not available in the other validation sets.

For platelet < 150.0 $10^3$/µL and WBC < 4.0 $10^3$/µL, the external eye model achieved AUCs of 59-73%, significantly higher than the baseline in validation set A (by 7.3% and 12.6%, respectively, $p < 0.0001$ for both), but not validation set B (these labels were not available in validation set C). For the TSH > 4.0 mU/L task, the external eye model had a 3.3% higher AUC than the baseline in validation set A, but this did not meet statistical significance ($p = 0.14$).

We also report positive predictive values with thresholds based on the 5% of patients with the highest predicted likelihood for each prediction target, and in each respective validation set (Supp. Table 2). Overall, we observe similar trends here, though p-values were generally higher due to the smaller denominator (only 5% are predicted as positive).

Supp. Table 5 summarizes results for all variables and cutoffs that we included in the model development. A few additional tasks performed well and merit future validation. For example, for the ALT > 29.0 U/L, BUN > 20.0 mg/dL, potassium > 5.0 mEq/L, sodium < 136.0 mEq/L tasks, the external eye model outperformed the baseline by 5.1%, 6.6%, 6.1%, 4.2%, respectively, in validation set A. Unfortunately, these variables were not available in the other validation sets. Appendix 6 discusses subgroup analysis for the primary analyses as well as analysis adjusting for multiple baseline variables, to account for possible correlation between variables. Appendix 7 further examines the image resolution requirements for the primary analyses.

**Explainability analysis**

To get insight into what parts of the image are most important to the DLS performance we conducted ablation experiments to mask different regions of the image during both training and evaluation, resulting in the DLS only seeing certain parts of the external eye. Masking involving the pupil and iris use location information predicted by an iris and pupil detector (see Appendix 3). In addition to this, we included a model that operated on grayscale images to assess the importance of color information. Cataracts served as a positive control in these experiments, as we would expect to see most of the signal originate in the pupil (i.e., performance should suffer if the pupil is masked and not visible). Additional details for this experiment are provided in Appendix 4. Figure 2 provides an example of each ablation and lists the AUCs of each model. Results for cataracts are as expected, with the performance being unaffected by removal of any portion other than the pupil, whereas removing the pupil (whether solely removing the pupil, or via masking all regions except the iris) decreased performance substantially. The results for other prediction tasks suggest that the signal is more distributed across the image and is not isolated to just the iris or pupil. In other words, removing the iris or pupil does not cause a great reduction in performance; correspondingly, a model trained on *only* the pupil or iris does not perform as well (but still generally performs better than the baseline). Color information appears to be somewhat important for most tasks, with Hgb < 11.0 g/dL, albumin < 3.5 g/dL, and calcium < 8.6 mg/dL being most affected by removal of color.

# Discussion

Our study demonstrates that a deep learning system can detect biomarkers of systemic disease, specifically kidney function and blood count abnormalities, from images of the external eye alone. We developed and evaluated this DLS using external eye image datasets obtained from diverse populations across the United States. The results showed that our DLS performed significantly better than baseline clinicodemographic models at predicting kidney function and blood count abnormalities across all three validation sets and at predicting abnormalities in liver and multiorgan parameters in validation set A. Results for several other systemic parameters appeared promising, and are discussed in Appendix 9. We next discuss our findings and their potential implications on an organ system by organ system basis.

For the detection of abnormalities in kidney function, a detailed review of the performance outcomes showed that the DLS performed best at more severe disease thresholds. For ACR, the model improvement compared to the baseline was more pronounced in increased severity of albuminuria (i.e. DLS improvement above the baseline was greatest for ACR > 1500 mg/g and least for ACR > 30 mg/g). Similarly, eGFR performance improvement was most substantial in more severe kidney disease. In datasets with a predominantly older population, more severe declines in kidney function were needed for the model improvement to become statistically significant as compared to the baseline. We theorize that since kidney function gradually declines with age[11], a markedly steeper drop in glomerular filtration rate may be required to manifest as an abnormality in an aging population.[12] Consistent with this hypothesis, we found that in younger subgroups, in which eGFR is usually highest, the model could detect more subtle deterioration of kidney function. These observations suggest that tools based on this DLS would thus be best utilized in screening settings, such as for detection of moderate kidney disease in a young, healthy population or for severe kidney dysfunction in an older population (both of which can be asymptomatic).[13,14]

With respect to decreased Hgb levels, non-invasive screening for anemia has long been considered with physical examination, such as findings of pale oral mucous, conjunctiva, or nailbed[15–17]. Numerous studies have shown that examination of the conjunctiva can detect severe anemia[15,18,19], thus, it is not surprising that our DLS had a statistically significant superior performance as compared to the baseline for detection of low hemoglobin. Our study is unique, however, in that images of the external eye were captured without any manipulation, such as pulling the eyelid down to expose the palpebral conjunctiva, which has been the more common site for evaluating anemia.[17,20] In fact, our ablation experiments (Figure 2) suggest that DLS is detecting other novel signals for anemia, not only from the color of the conjunctiva, since our model still outperforms the baseline when the external eye image is in grayscale and when all eye structures are masked except for the iris.

An unexpected outcome from our study was that despite performing better than the baseline, the absolute performance for liver (AST) and thyroid (TSH) abnormalities was lackluster, with AUCs in the low 60's. In a recent study[21], Xiao et al developed a DLS for the detection of hepatobiliary disease from slit lamp and fundus photos. They focused on chronic liver conditions and found that relative to milder disease, their DLS had improved performance in cases of advanced liver disease (cirrhosis, liver cancer). Additional studies on chronic liver disease and more severe disease will be needed to assess the performance of our DLS. Their model's performance was also better for slit lamp photos, which capture images of the external eye and anterior segment, as compared to fundus photos, which capture images of the retina. This is in-turn consistent with previous literature that, in addition to icterus, other external eye findings, such as conjunctival and corneal xerosis, Bitot's spot, keratomalacia, dry eyes, and corneal ulcers, are known to be associated with liver disease[22] in part due to vitamin A and D deficiencies.[23–25] Similarly, thyroid disease is known to be associated with numerous external eye findings, such as conjunctival hyperemia and chemosis, lid retraction, and proptosis[26]. We suspect that the performance for thyroid disease was largely related to the low number of positive cases for the selected thresholds in our datasets, and larger studies using more severe thresholds may yield better results.

On the topic of the DLS's potential uses, our analysis on image resolution sensitivity showed that even with low resolution images (75 pixels across, corresponding to less than 1% of the pixel count of modern smartphone cameras), the DLS still outperformed the baseline for several systemic parameters. This promising observation is in line with prior diabetes-related external eye research[6].

Taking into consideration both the low image resolution requirement and lack of need for extensive clinical training, external eye photography (such as via smartphones) may be easier to use by the general user population compared to clinician-focused tools such as a slit lamp. Across the different predictions where the DLS outperformed the baseline, the AUCs ranged from 61.5% to 87.7%, which may not be accurate enough for a diagnostic device, but is in line with other scenarios like cardiovascular risk assessment[27], mammography[28] and pre-screening for diabetes[29].

Several limitations apply to our study. First, with respect to population characteristics, our validation sets were primarily from a diabetic retinopathy screening population. The exception was validation set B, which was from a general eye screening program, albeit the subset of this dataset that had systemic labs available was again (and perhaps unsurprisingly) predominantly diabetic. Second, regarding cameras used to photograph the external eye, all images were collected on fundus cameras and it is yet unknown if images collected via alternative camera types, such as a smartphone camera, would result in comparable DLS performance. Third, several subgroup analyses showed a trend towards poorer performance that was difficult to interpret in light of limited sizes of subgroups (e.g. age, sex, BMI, race; see Appendix 6). More focused data collection for DLS refinement and evaluation across subgroups will be needed before considering clinical use. In addition, further development and evaluation on future datasets where cystatin C is available consistently will be important to ensure more accurate eGFR measurements, particularly for Black patients[8]. Thus, further studies are needed to determine generalizability to broader patient populations and alternate devices.  Fourth, additional variables like medication and comorbidities were not consistently available and therefore could not be incorporated into the baseline models. Lastly, insights from our explainability experiments were limited to the location of the signal, for example, cornea/iris and pupil. Additional explainability research may improve our understanding of the specific features learned by the DLS and whether signs of these diseases are visible to the human eye.

Leveraging deep learning tools to detect systemic disease may be useful in several ways: individuals identified as having early or mild disease could receive early intervention to prevent progression, whereas individuals detected to have more severe disease may be prioritized for immediate care. Previous work has explored other modalities for detection of systemic disease by developing DLS that utilize retinal fundus images[30], echocardiograms[31], and electrocardiograms[32,33] to identify systemic disease. These studies have required specialized equipment or medication, such as dilating drops, which are not readily available outside of medical facilities. Our study suggests that non-invasive imaging of the external eye has the potential to provide information about systemic disease without the use of specialized equipment or medication. Removing these barriers to clinical assessment may have substantial health impact. Further studies are needed to determine if this DLS could identify systemic disease markers from external eye images captured by other camera types and how external eye screening could be effectively adapted and utilized in both clinical and non-clinical settings.

# Contributors

BB, CC, NH, and YL conceived and designed the study. BB and YL analyzed the data. IT conducted the literature review. JC, LPD, AYM, AV, PS, and NH were involved in data collection and curation. AU, RK, EK, YM, GSC, LP, DRW, CS, JK, and AVV provided strategic guidance and oversight. GSC, LP, DRW, CS, JK, and AVV obtained funding. BB, IT, NH, and YL drafted the manuscript with feedback from all authors. BB and YL verified the data. All authors contributed to data interpretation and read and approved the final manuscript.

# Data sharing

This study utilized de-identified data from EyePACS Inc., and the Technology-based Eye Care Services and the teleretinal diabetes screening program at the Atlanta Veterans Healthcare System. Interested researchers should contact J.C. (jcuadros@eyepacs.com) to inquire about access to EyePACS data and approach the Office of Research and Development at https://www.research.va.gov/resources/ORD_Admin/ord_contacts.cfm to inquire about access to VA data. Those interested in retraining a model can find the pretrained (BiT-M) architecture at https://github.com/google-research/big_transfer/blob/master/README.md

# Declaration of interests

BB, CC, PS, AU, YM, GSC, LP, DRW, CS, JK, AVV, NH, and YL are Google employees and own Alphabet stock. IT is a paid consultant to Google. All other authors declare no competing interests.

# Acknowledgements

The authors would like to acknowledge Drs. Dave Steiner, Yuan Liu, and Michael Howell for manuscript feedback; Elvia Figueroa and the LAC DHS TDRS program staff for data collection and program support; Andrea Limon and Nikhil Kookkiri for EyePACS data collection and support; Dr. Charles Demosthenes for extracting the data and Peter Kuzmak for getting images for the VA data. Part of the data processing team at LAC DHS was supported by grants UL1TR001855 and UL1TR000130 from the National Center for Advancing Translational Science (NCATS) of the U.S. National Institutes of Health. The content is solely the responsibility of the authors and does not necessarily represent the official views of the National Institutes of Health, Department of Veteran Affairs or of the US Government.

# Tables & Figures

**Table 1. Dataset characteristics**

Baseline variables (age, sex, race/ethnicity, and years with diabetes) were not available for all patients. For categorical variables below, we specify the number of patients per variable value and the percentage of *available* data.

| Dataset | Development sets | | Validation set A | Validation set B | Validation set C |
|---|---|---|---|---|---|
| Source | EyePACS /LACDHS (Train) | EyePACS /LACDHS (Tune) | EyePACS /LACDHS (Test) | VA TECS | VA TRI |
| Geographical location | California | California (independent sites from Train/Test) | California (independent sites from Train/Tune) | Georgia | Georgia |
| Number of patients | 40,427 | 8,598 | 12,205 | 10,402 | 6,266 |
| Number of visits | 62,213 | 14,245 | 17,363 | 13,348 | 6,485 |
| Number of images | 123,130 | 28,137 | 34,159 | 24,662 | 12,616 |
| Age (years, mean ± std. dev.) | 56.4 ± 10.3 | 56.7 ± 9.7 | 56.5 ± 10.0 | 63.3 ± 9.9 | 61.9 ± 12.8 |
| Female | 24,322 (60%) | 5,425 (63%) | 6,741 (55%) | 478 (5%) | 944 (15%) |
| Male | 16,095 (40%) | 3,172 (37%) | 5,460 (45%) | 9,924 (95%) | 5,322 (85%) |
| Hispanic | 26,494 (75%) | 5,209 (70%) | 8,634 (80%) | 0 (0%) | 0 (0%) |
| White | 1,867 (5%) | 512 (7%) | 397 (4%) | 2,323 (52%) | 2,577 (43%) |
| Black | 3,835 (11%) | 668 (9%) | 645 (6%) | 2,089 (47%) | 3,317 (56%) |
| Asian / Pacific islander | 2,514 (7%) | 939 (13%) | 465 (4%) | 29 (1%) | 38 (1%) |
| Native American | 36 (0%) | 27 (0%) | 4 (0%) | 21 (0%) | 36 (1%) |
| Other | 531 (2%) | 80 (1%) | 581 (5%) | 0 (0%) | 0 (0%) |
| Years with diabetes (years, median (inter quartile range)) | 8.0 (2.0 - 13.0) | 8.0 (3.0 - 13.0) | 8.0 (2.0 - 13.0) | N/A | N/A |

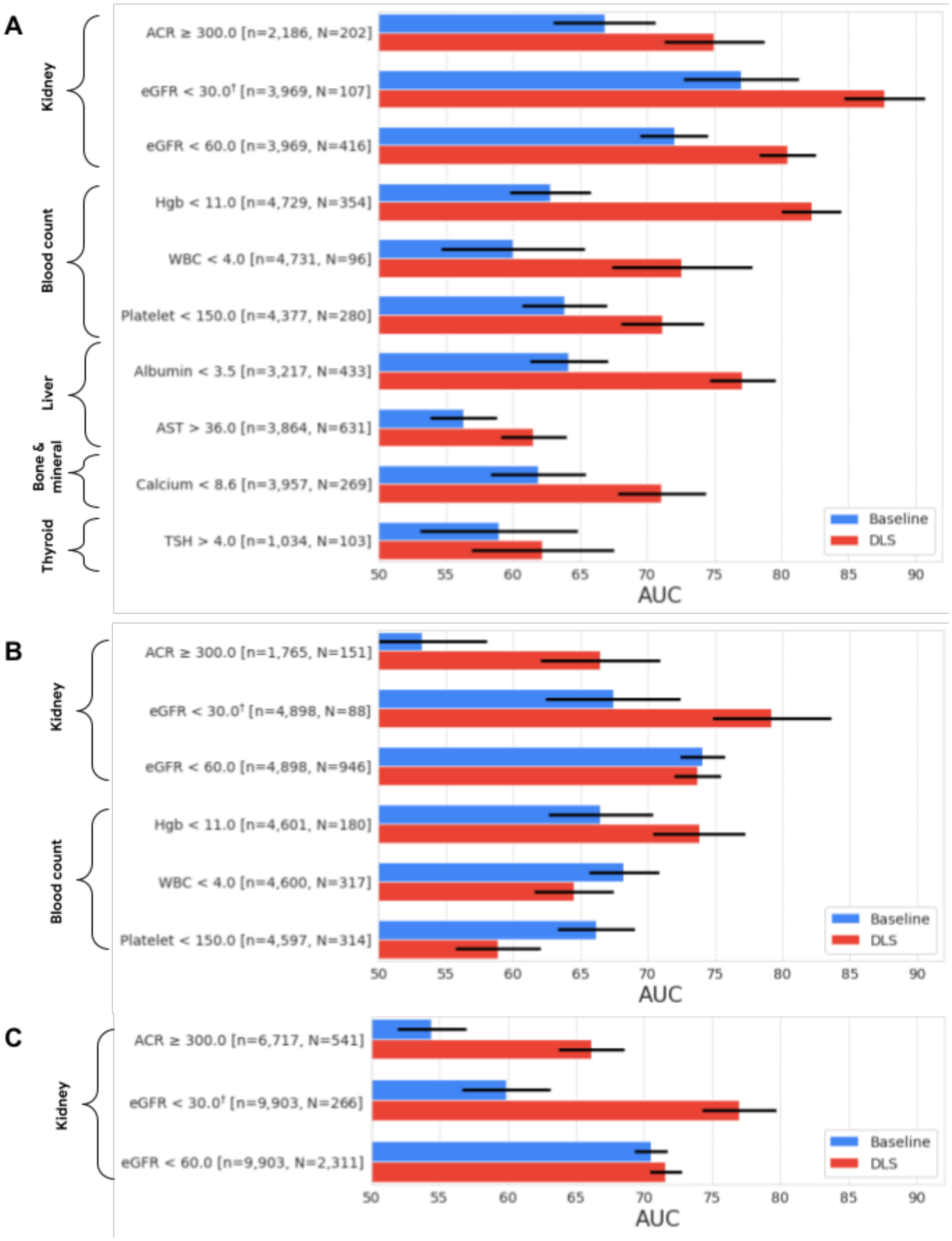

**Figure 1. Comparison of AUC of the baseline and the DLS**

Results are presented for validation sets A, B, and C. Error bars show 95% confidence intervals computed using the DeLong method. Supp. Table 1 contains these results in tabular format, and includes p-values. †Indicates that the target was prespecified as secondary analysis; all others were prespecified as primary analysis.

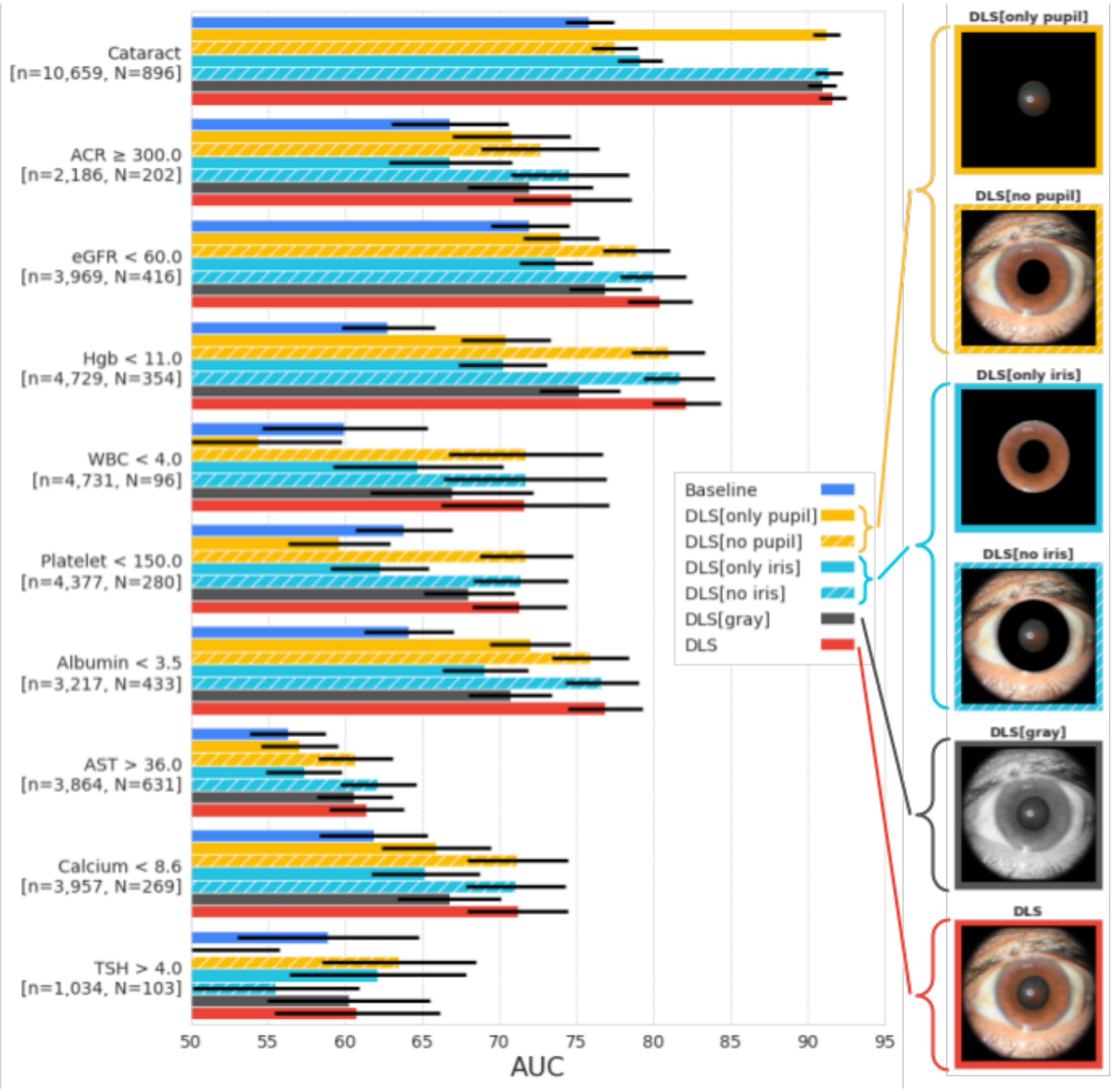

**Figure 2. Experiments masking different image regions or removing color**

Results are presented for validation set A. The reported performance are for models both trained and evaluated on images with the specified manipulations. Manipulations described as "only X" refer to images with only that part of the anatomy visible; "no X" refers to images with that region masked, but every other area visible; "gray" refers to images converted to grayscale to remove color. Error bars show 95% confidence intervals computed using the DeLong method. For details, see Appendix 4. Note that all DLS results here are single models rather than ensembles.

# Supplementary Material

## Appendix 1. Additional modeling details

We included several additional "auxiliary heads" to boost performance: age discretized into 7 buckets, race/ethnicity, self-reported sex, years with diabetes discretized into 5 buckets, diabetic retinopathy (DR) grade, presence of diabetic macular edema (DME), presence of vision-threatening DR (VTDR), and whether the patient is a self-reported smoker. Finally, we included the presence of cataracts as an additional prediction to serve as a positive control. Note that not all labels were available for every photo, and hence loss was only propagated for heads for which a label was available.

We found that using a ResNet BiT-M architecture, pre-trained on ImageNet-21k[34,35] produced superior results on our tune set over the Inception-v3[36] architecture used in Babenko, Mitani, et al[6]. We found further performance gains by first training a model on the tasks and dataset reported in that prior work, and then performing additional fine-tuning on our training dataset (see Supp. Figure 3; note that while our prior work also used data from EyePACS to train the model, we excluded the LAC area in those experiments to ensure independence of the EyePACS data used for this study). The effects of some of these modeling choices are discussed in the next section.

To select hyperparameters, we ran a search with 50 random trials sweeping over the parameters listed in Supp. Table 7 (B). For early stopping, we used the tune-set area under the receiver operator characteristic curve (AUC) of the eGFR < 60.0 mL/min/1.73 $m^2$ head. Using the same criteria, we selected the best 5 models for our final ensemble.

## Appendix 2. Effect of modeling choices

Supp. Figure 3 demonstrates the effect of a few modeling choices. Specifically, we found little difference between the 5-model ensemble and a single model, and we found that fine-tuning from the model trained on the dataset reported in Babenko, Mitani, et al[6] provided a modest, but consistent boost in performance across the various tasks.

Finally, we evaluated the performance of a DLS model that takes both an external eye image and the metadata variables in the baseline as input. We evaluated this model on validation set A, and found that it did not outperform the DLS model that takes only the external eye image as input.

## Appendix 3. Pupil and iris detection

To determine whether pupil size is a confounder and perform the ablation experiments described in the next section, we leveraged the pupil and iris detection system presented in our prior work. For details we refer the reader to that paper, but briefly, the detection system returns the parameters of two axis-aligned ellipses, one for the iris and one for the pupil. Because the precise distance from the eye to the camera is not known for each image, we exploited the fact that iris (or cornea) size is roughly constant across individuals[37], and computed a normalized pupil size (which we refer to as simply "pupil size" throughout) by dividing the pupil size in pixels by the iris size in pixels (where both of these "sizes" are the averages of height and width of the respective ellipses).

**Appendix 4. Ablation analysis**

To assess which parts of the external eye photos were most important for the various predictions, we performed ablation experiments for image regions and image color. Specifically for the image region ablation, we used the pupil and iris detection system described above to either mask out the pixels *within* the pupil or iris (i.e. erase these pixels and set them to black), or mask out all pixels *outside* these regions. We then trained a model on the resulting images, and measured performance on validation set A with the same masking applied. Note that while these experiments withhold information contained *within* various parts of the eye, the *size* of each ablated region (e.g. pupil size) can still be derived by the model from the size of the masked regions. For the image color ablations, color was removed by converting the images into grayscale using the tf.image.rgb_to_grayscale method in the TensorFlow library, which applies the following weights: 0.2989 for R, 0.5870 for G, and 0.1140 for B. For modeling convenience and to ensure parity in the number of model parameters, these channels were replicated such that each image contained 3 channels with identical (grayscale) information.

As discussed in the main text, the results suggest that the information is distributed across the external eye photograph, instead of being isolated to the pupil or iris. Additional ablation experiments involving anatomies beyond the pupil and iris may be useful to further understand what areas of the image are the most informative, and correspondingly, how to optimize future data collection efforts.

**Appendix 5. Temporal sensitivity analysis**

All experiments discussed thus far used the strategy described in the Methods for matching labs and vitals measurements to a given photo, by filtering labs using the time difference between the two. To see if our results were sensitive to the choice of this time difference cutoff we evaluated the model on subsets of the data with stricter cutoffs. Supp. Table 4 shows results. We note that there was little difference across the board, with all the AUCs for stricter cut-offs being within 3% of the original results (with the exception of TSH > 4.0, which had only 31 positive examples after stricter filtering).

**Appendix 6. Subgroup and adjusted analysis**

We performed extensive subgroup and adjusted analysis on a number of variables in all 3 validation sets (Supp. Table 3, Supp. Table 6). As in our primary analysis, for each subgroup we compute the AUC improvement of the DLS above the baseline, and compare this against the improvement in the overall dataset. Generally, we observed that loss of statistical significance of the improvement happened in subgroups with relatively few (e.g., < 50) positive examples.

For the eGFR < 60.0 target, we observe that in validation sets A and C, both of which have diabetic populations, there is a strong trend between age and the AUC improvement: the external eye model performs worse in older populations. This trend does not appear in validation set B, though that population consists of both patients with and patients without diabetes. We also note that there are some differences in performance across race/ethnicity groups, but the sample sizes (particularly for the eGFR < 30.0 target, for which the model performed better) were quite small.

For ACR ≥ 300 (which is another marker of kidney function), in validation sets B and C, where most patients were Black or White, the AUC improvements for both subgroups are within 5% of the improvement on the full set. This analysis was not possible to do reliably in validation set A, where approximately 70% were Hispanic.

For the Hgb < 11.0 target, AUC improvement was within 1% for male and female subgroups, compared to improvement in the overall dataset. We note that performance was worse in the non-diabetic subgroup in validation set B, though the DLS still outperformed the baseline. Compared to the eGFR analysis, we also observed the opposite trend with respect to age: performance appeared poorer in the younger groups in validation sets A and B (Hgb was not available in set C). Performance also trended poorer for Black patients in validation set B in particular.

Given above, we additionally examined eGFR and Hgb prediction performance adjusted for the baseline variables (Supp. Table 6), and found that the DLS's predictions remained highly significant (p < 0.0001) after adjusting age, sex, race/ethnicity, and years with diabetes (where available) in all 3 validation sets. Taken together, the subgroup and adjusted analyses suggest that the DLS's performance for kidney function and blood count (Hgb, WBC, platelets) appeared to vary across subgroups, but the mechanism of variation does not appear to be due to the DLS relying on predicting other baseline variables (e.g., identifying patient age, sex, or race). Given the relatively small number of positive examples in subgroups, these trends and observations will need further study in additional cohorts. For Hgb, it does not appear implausible that the DLS is quantifying pallor, and that the strength of this signal is attenuated by skin or iris pigmentation. Further careful study is needed before considering the use of such a model clinically.

**Appendix 7. Image resolution sensitivity analysis**

Similar to our previous work, we conducted an experiment to assess the impact of image resolution for each prediction task. We trained and evaluated models with images that were downsampled using the area-based method (tf.image.resize with method=AREA). To keep the model architecture and parameter count unchanged, after downsampling the images, we upsampled them back to the resolution used in the original models (by bilinear interpolation with antialias using tf.image.resize with antialias=True). Supp. Figure 1 (A) illustrates what each downsampled image looks like, and (B) shows how performance on validation set A degrades as the resolution is reduced. For all prediction tasks performance appears stable down to 150 by 150 pixels, and then begins to degrade. In most cases, even at 75 by 75 pixels, performance of the DLS is still better than the baseline. The most surprising result here is that of Hgb < 11.0: even at a resolution of 5x5, the DLS significantly outperforms the baseline, suggesting that a lot of information comes from broad color information (this is corroborated by the ablation studies discussed in the previous section). Supp. Figure 1 (C) shows similar plots for validation set B. Here we see that performance of Hgb < 11.0 degrades more quickly than in validation set A, suggesting that the signal from color information alone may not easily generalize to different capture protocols or environments (and indeed, qualitatively we have observed that the images in validation set A have a different color balance compared to the ones in validation set B, where the entire image tend to have a yellowish tint).

**Appendix 8. Predictive power of additional variables**

To assess whether additional variables would improve baseline performance, we trained and evaluated additional baseline models that include systolic and diastolic blood pressure (BP), and body mass index (BMI). Note that because these variables were not available for all visits, both training and evaluation were done on a smaller amount of data than results reported elsewhere in this work. Supp. Figure 4 shows the results of these experiments for validation sets A and B. In validation set A, including BMI and BP in the baseline improves baseline performance by < 5% AUC in all but one prediction target (ACR ≥ 300.0), though even then the DLS's performance remains higher than this augmented baseline. In validation set B, including BMI in the baseline improves the baseline by < 5% AUC in all cases. Notably, for both validation sets, the addition of BP and/or BMI sometimes decreases performance, potentially indicating generalization challenges even with a simple baseline model.

In addition to these demographic variables, we also measured the predictive power of the pupil size, which was derived from the pupil and iris detector (see Methods). Supp. Figure 5 shows the AUC of the pupil size alone, and in combination with the rest of the baseline variables for validation sets A and B. The pupil size had modest predictive power for most prediction tasks (AUC in low 60s or lower), but when added to the other baseline variables, performance improved by less than 3% across all predictions.

Finally, Supp. Figure 6 shows the AUC of each individual baseline variable. Of note in validation set A, years with diabetes appeared to be a major contributor to the baseline AUC. Unfortunately the other validation sets did not have this variable, but future research should aim at including this information if possible.

**Appendix 9. Additional promising targets**

In our analysis, we chose to pre-specify a total of 9 prediction targets to reduce the chance of false positive findings. To supplement this, we provide the complete list of results in Supp. Table 5, which includes other potentially promising targets that may be worth exploring in the future. For example, BUN > 20.0, total bilirubin > 1.0, potassium > 5.0, HCT < 39.0 and HDL > 60.0 all had an AUC delta between DLS and baseline exceeding 5%; the last two were also available in validation sets B and C and had positive AUC deltas. Results for glycated hemoglobin and lipids (total cholesterol and triglycerides) are also provided and were similar to the prior study.

**Appendix 10. Detailed imaging protocols**

At LACDHS sites, the imaging protocol involved positioning patients in front of the camera, with their chin on the chin rest and their forehead approximately an inch away from the forehead brace. Both eyes were imaged while ensuring image clarity, focus, and visibility of the entire eye[38]. Topcon (NW8 and NW400) cameras were used. At the VA sites, external eye photographs were taken to document findings such as cataracts and eyelid lesions. The imaging protocol involved slightly distancing the patient from the camera, with further alignment entrusted to the operator. Topcon (NW8 and NW400) and Zeiss Cirrus Photo 600 cameras were used.

# Supplementary Figures

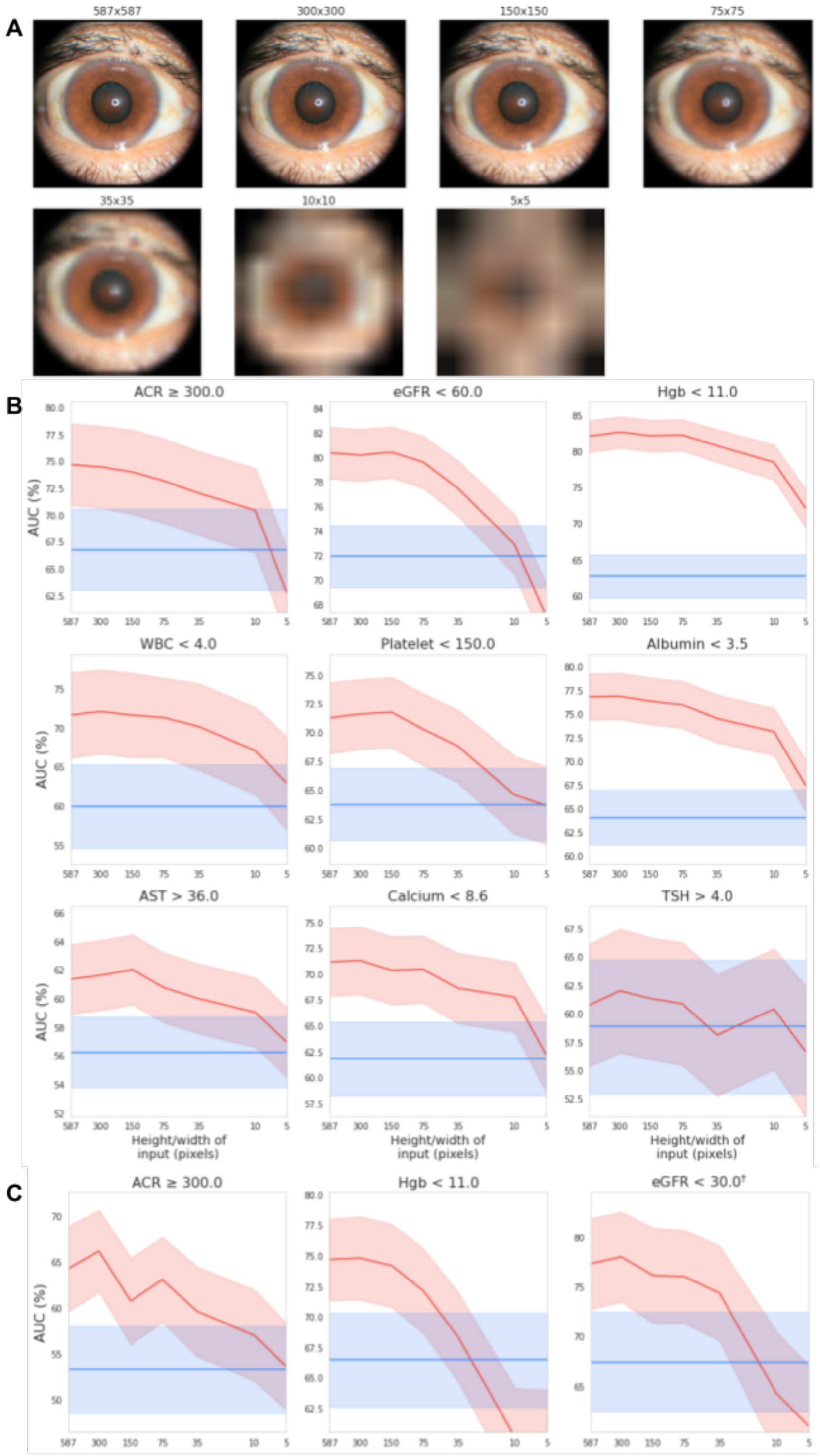

**Supplementary Figure 1. Effect of input image resolution**

(A) demonstrates a sample image scaled to different sizes used in this experiment. (B) and (C) show performance of models trained on different image sizes for validation sets A and B, respectively. Shaded region shows 95% confidence intervals computed using the DeLong method. See Appendix 7 for details**.** Note that all DLS results here are single models rather than ensembles. †Indicates that the target was prespecified as secondary analysis; all others were prespecified as primary analysis.

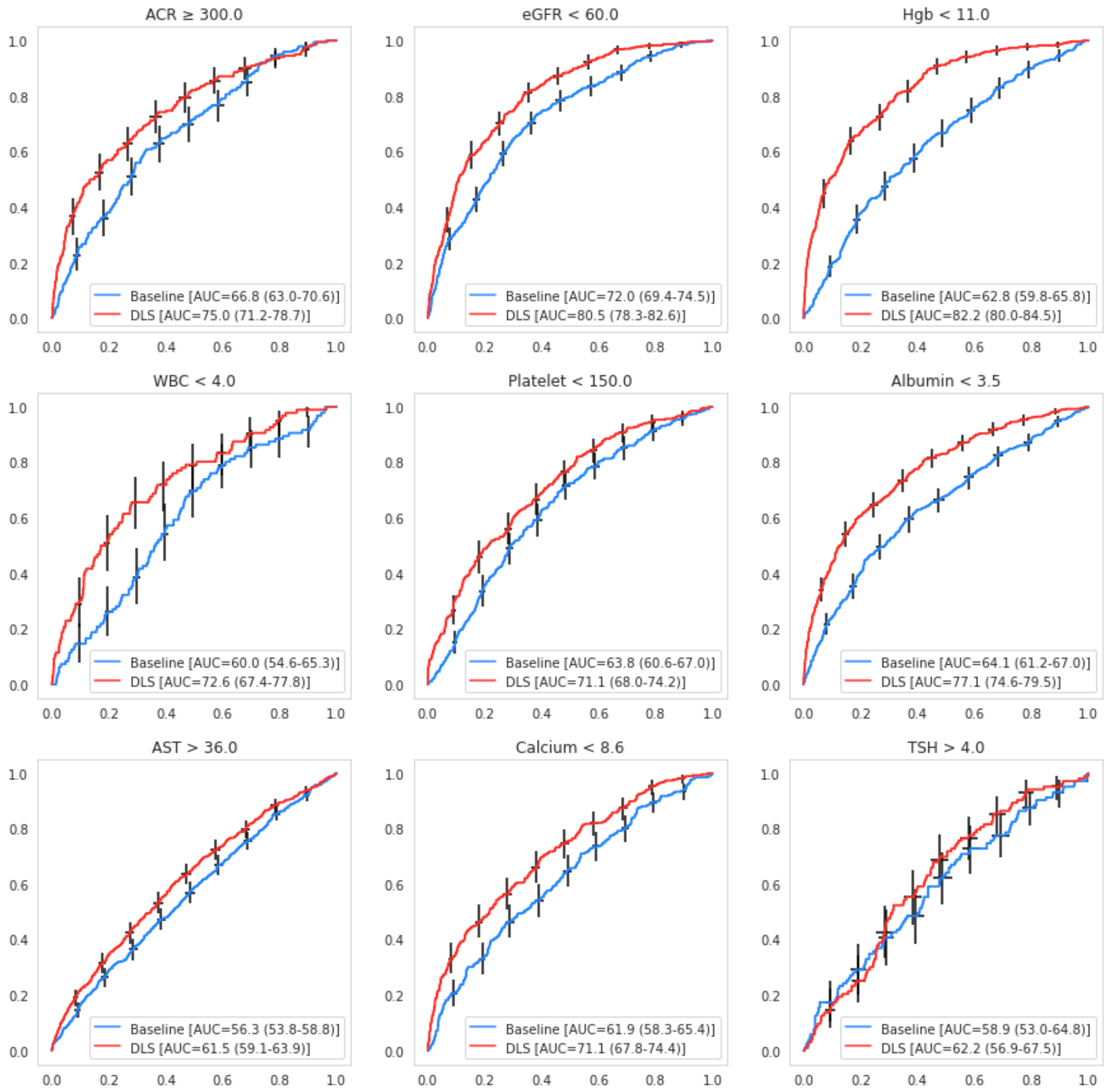

**Supplementary Figure 2. ROC curves per prediction, comparing DLS and baseline**

Results are presented for validation set A. Error bars show 95% confidence intervals for specificity and sensitivity, computed using bootstrap.

A

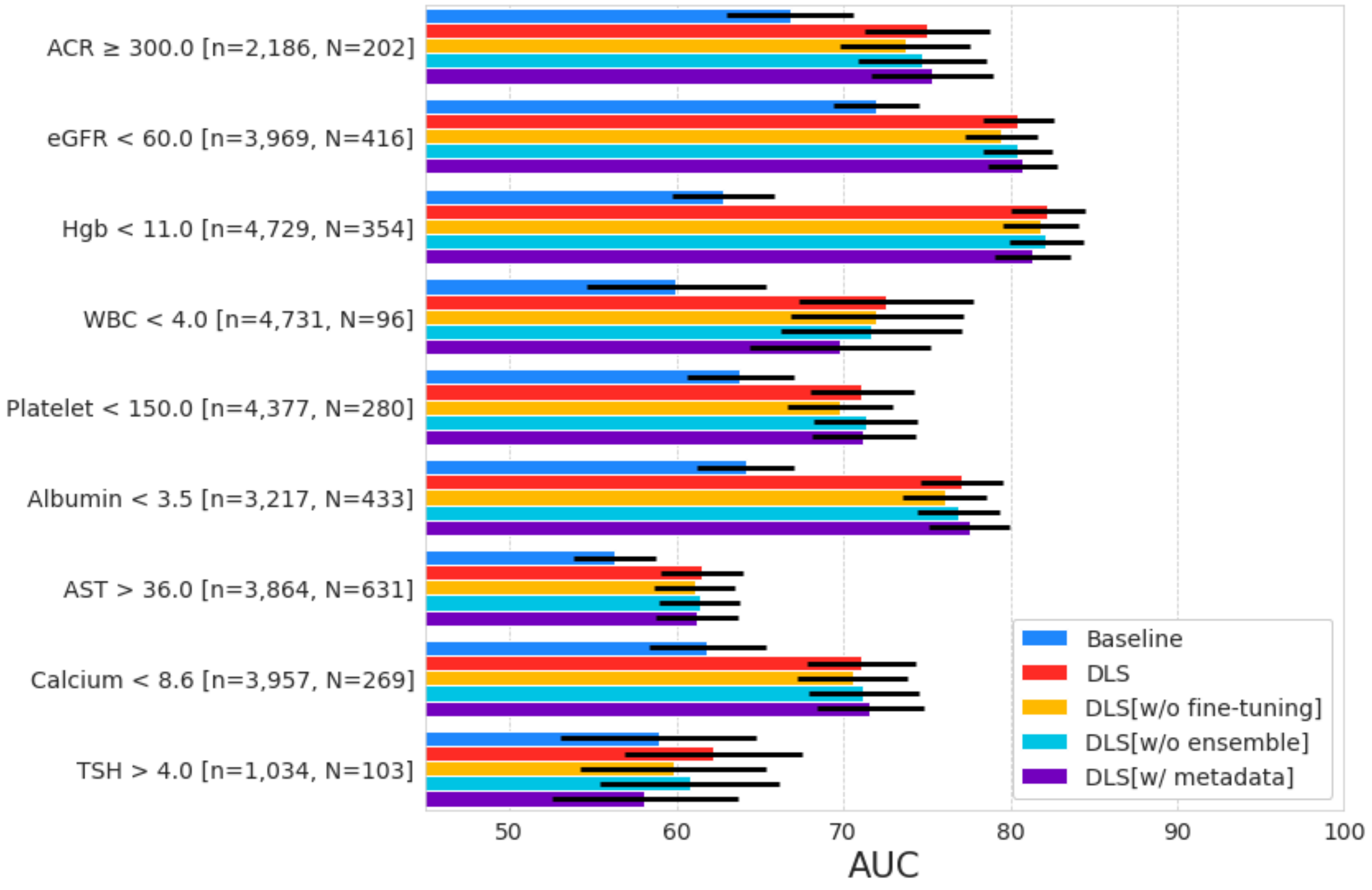

B

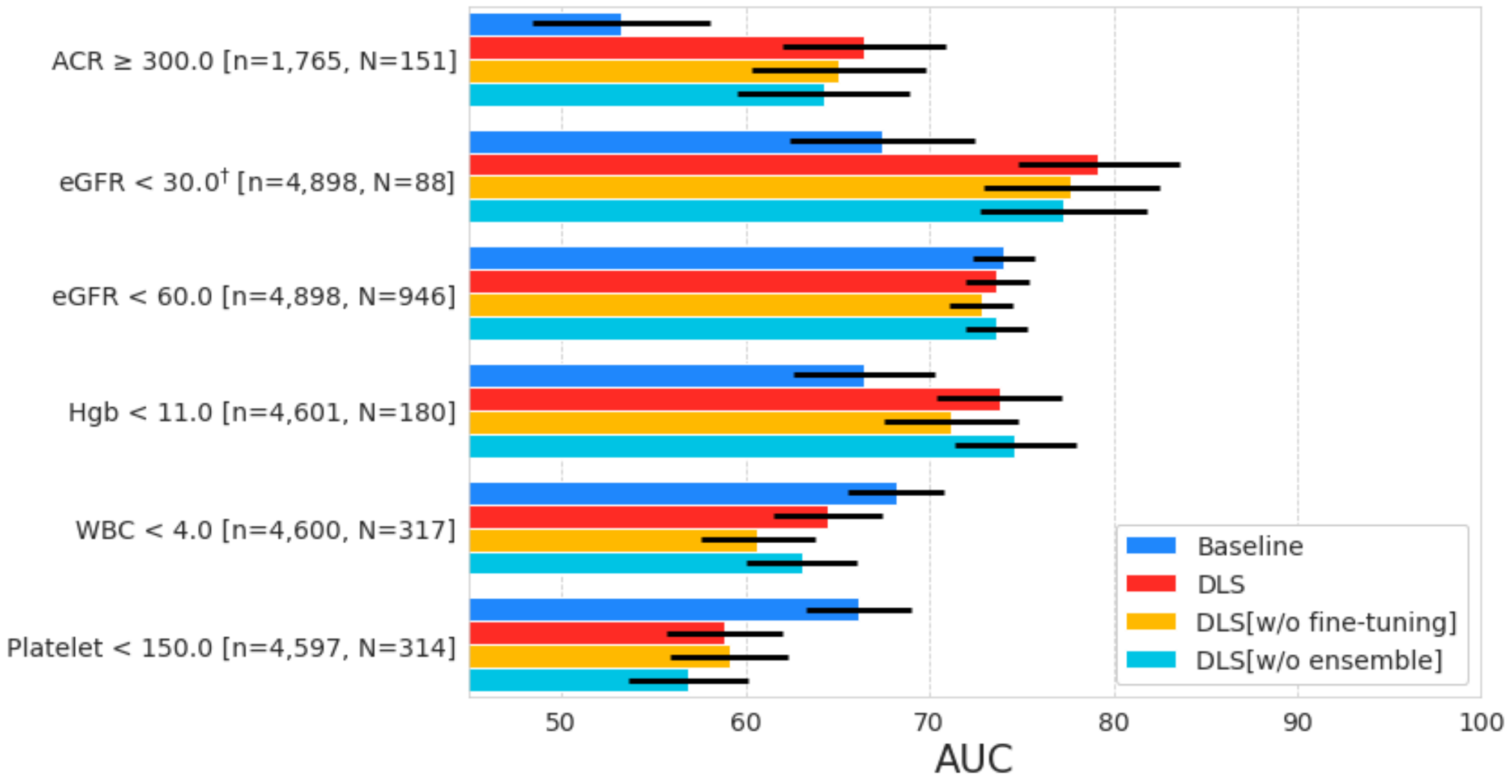

**Supplementary Figure 3. Comparing different DLS models**

Comparing the default DLS ("DLS"), DLS that was not fine-tuned from the model from our previous work (i.e. was finetuned directly from the BiT-M checkpoint trained on the ImageNet-21k dataset;

“DLS[w/o fine-tuning]”), a single DLS model (i.e. without ensembling; “DLS[w/o ensemble]”), and a DLS model that included metadata as input (“DLS[w/ metadata]”). (A) shows results for the validation set A, and (B) shows results for validation set B (excluding DLS[w/ metadata] because the available metadata variables differed in this dataset). Error bars show 95% confidence intervals computed using the DeLong method. †Indicates that the target was prespecified as secondary analysis; all others were prespecified as primary analysis.

A

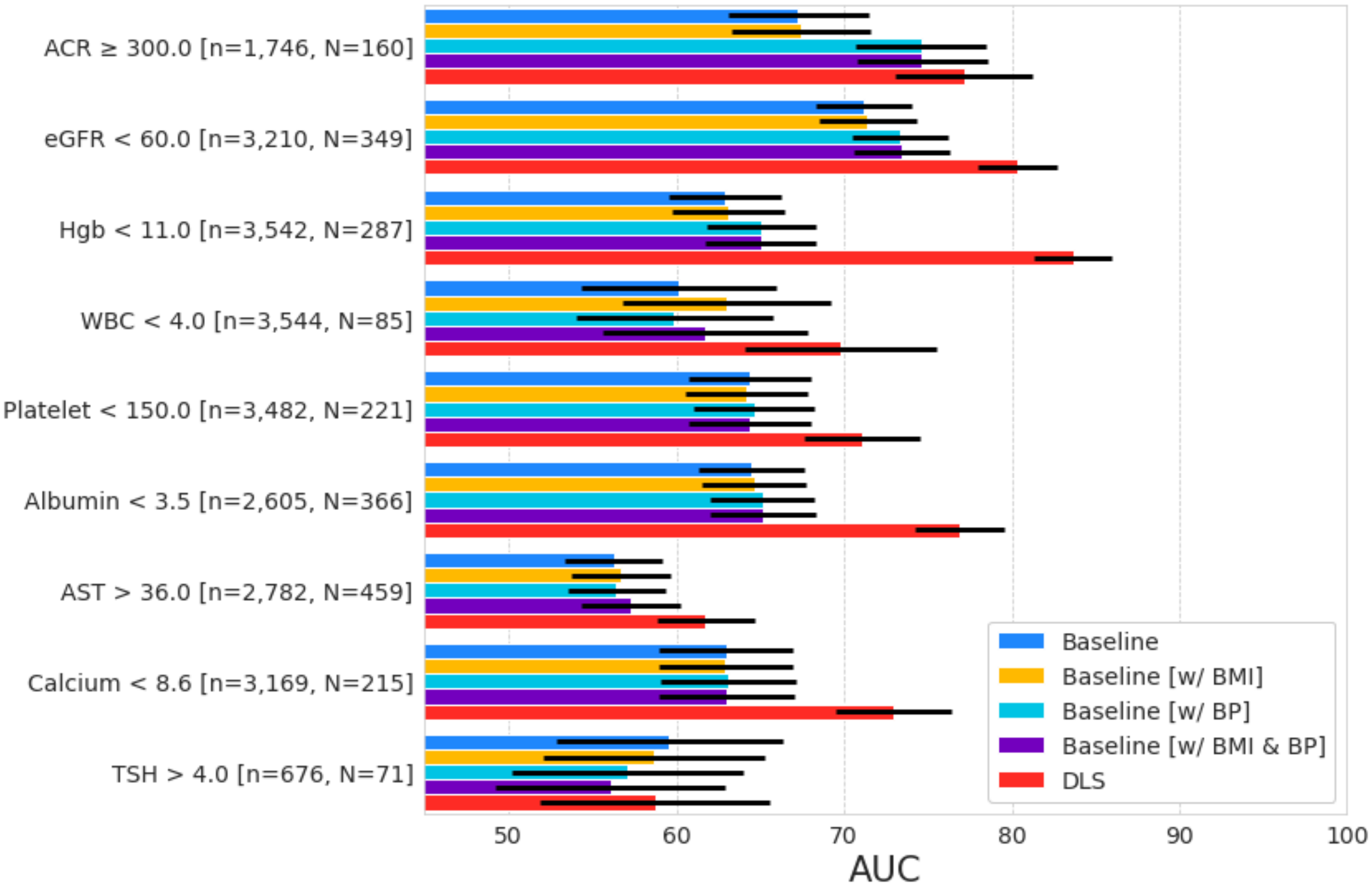

B

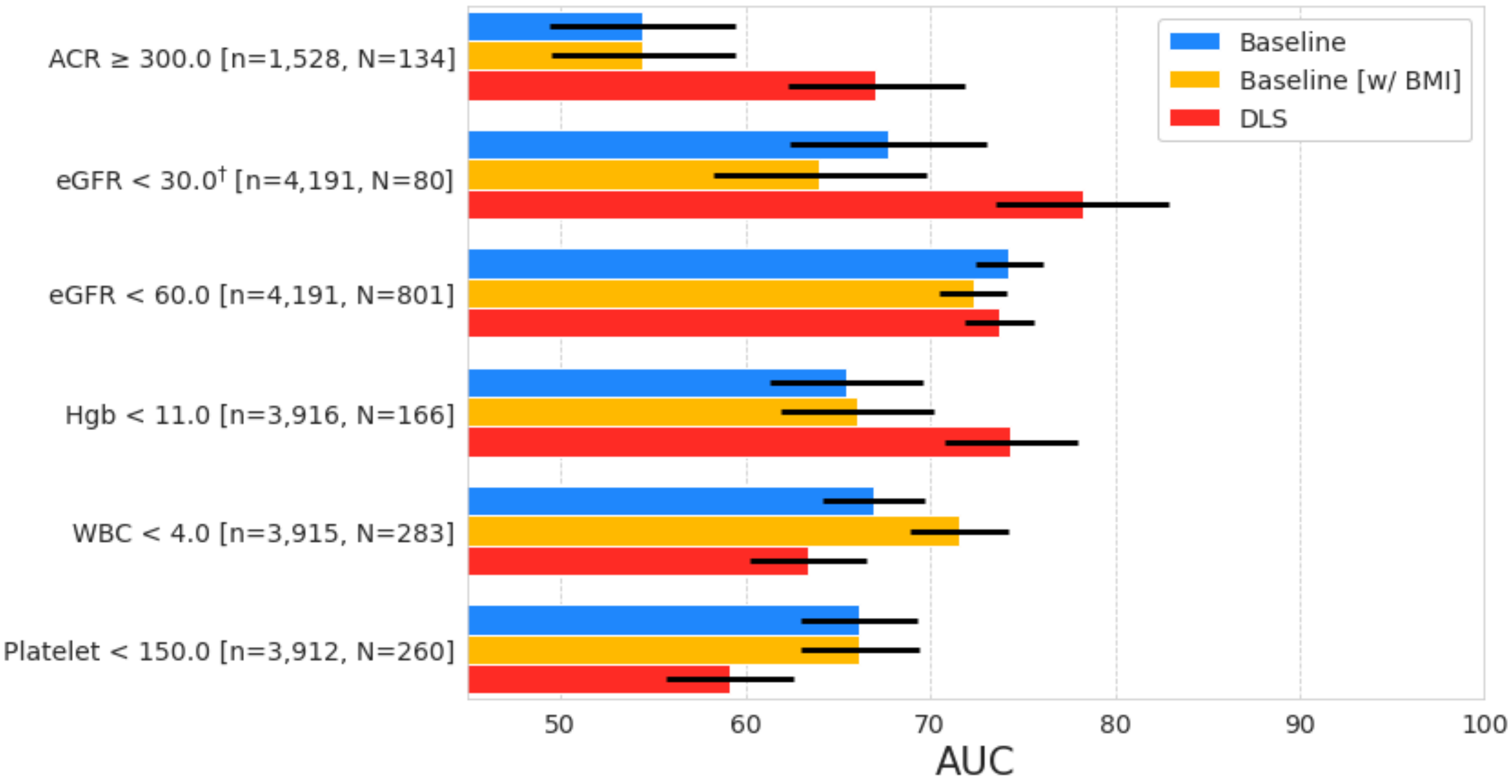

**Supplementary Figure 4. Comparing different baseline models**

Comparing the AUC of DLS and baseline models that include additional variables on subsets of data where all of these variables are available. (A) For validation set A, we include blood pressure (BP) and body mass index (BMI). (B) For validation set B, we include BMI. Error bars show 95% confidence intervals computed using the DeLong method.

A

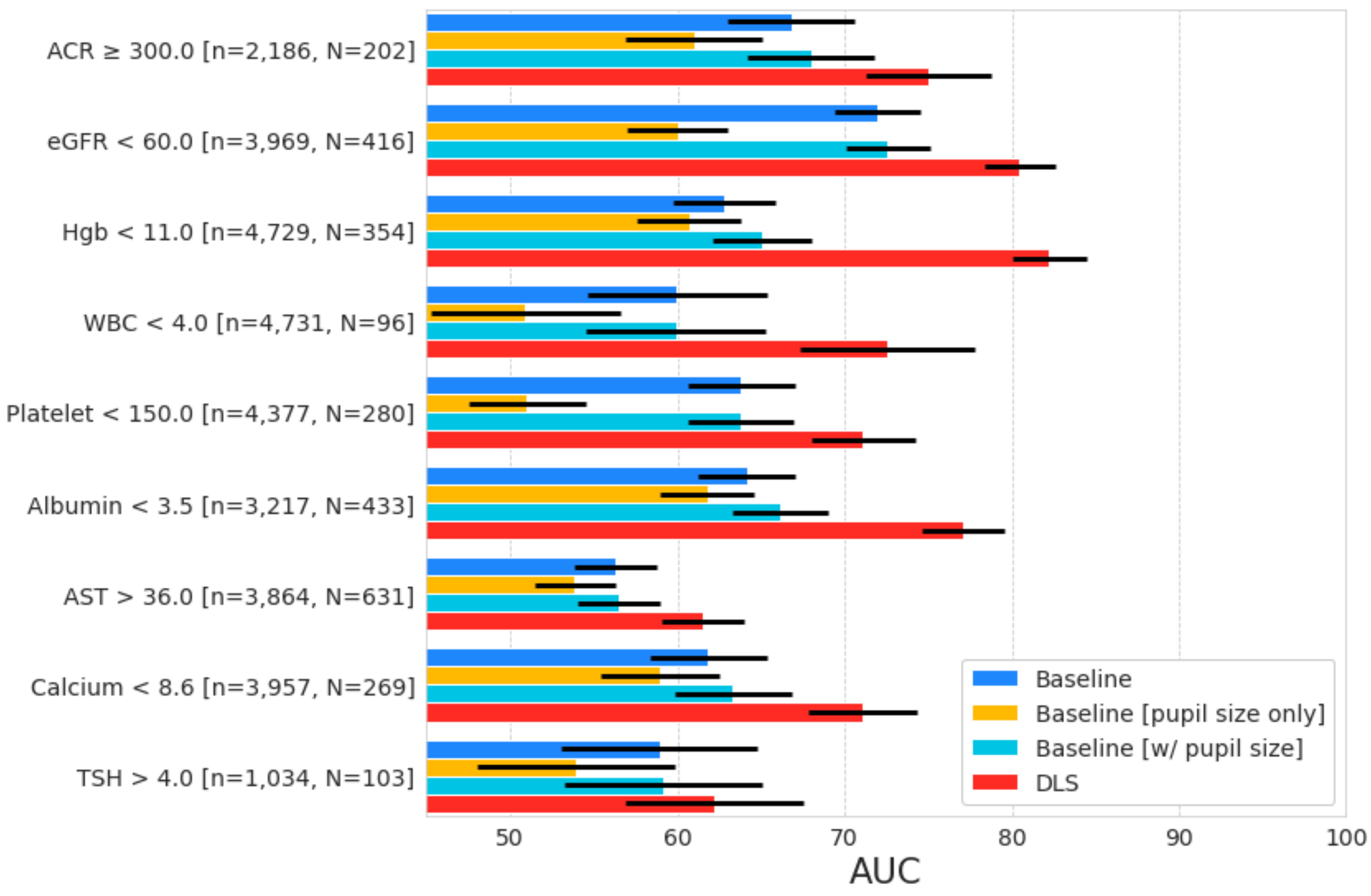

B

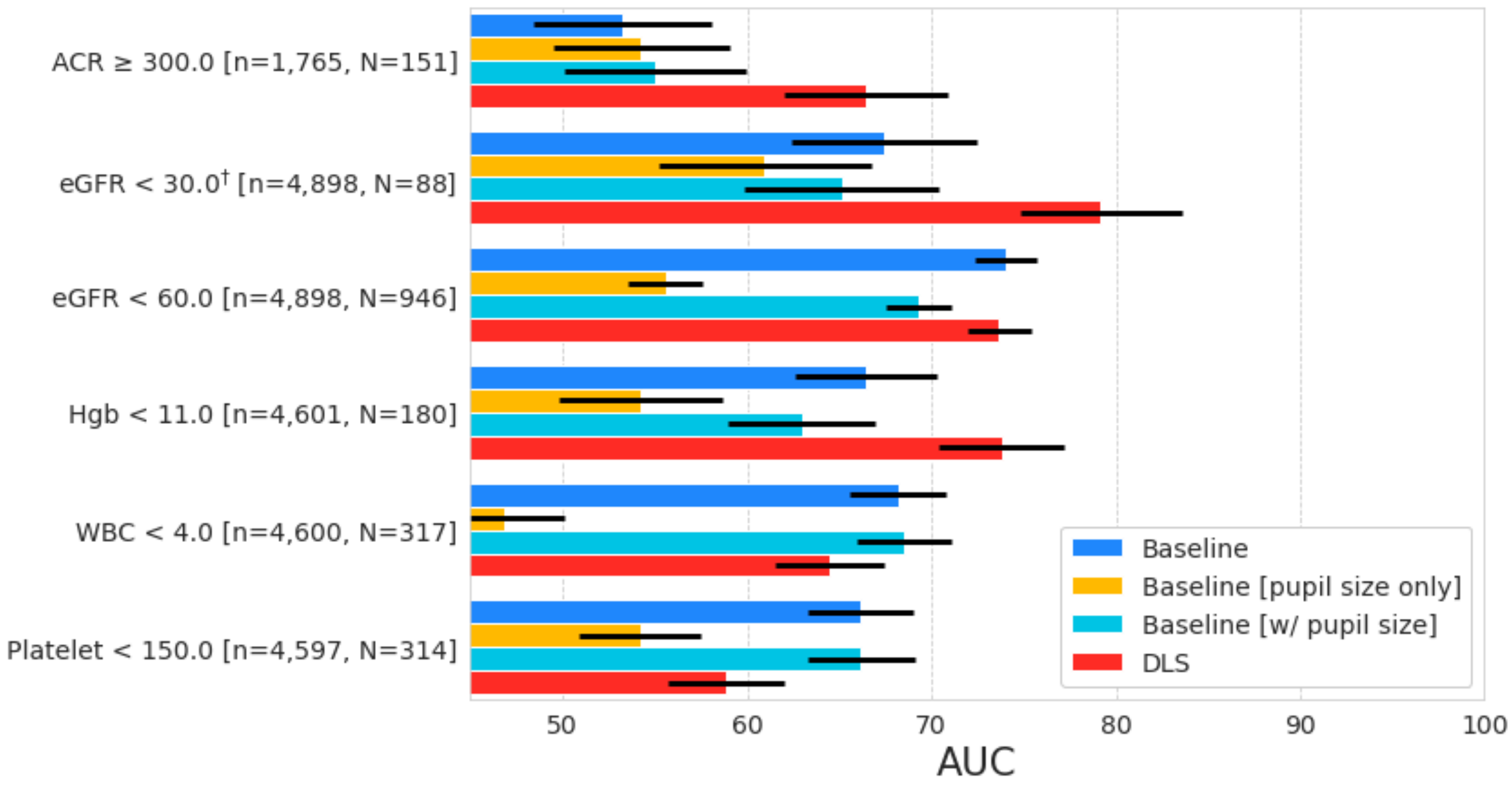

**Supplementary Figure 5. Predictive power of pupil size**

Comparing the AUC of the DLS, pupil size (as measured by our pupil and iris detection model), and a baseline model that includes the pupil size as an additional input for (A) validation set A and (B) validation set B. Error bars show 95% confidence intervals computed using the DeLong method.

A

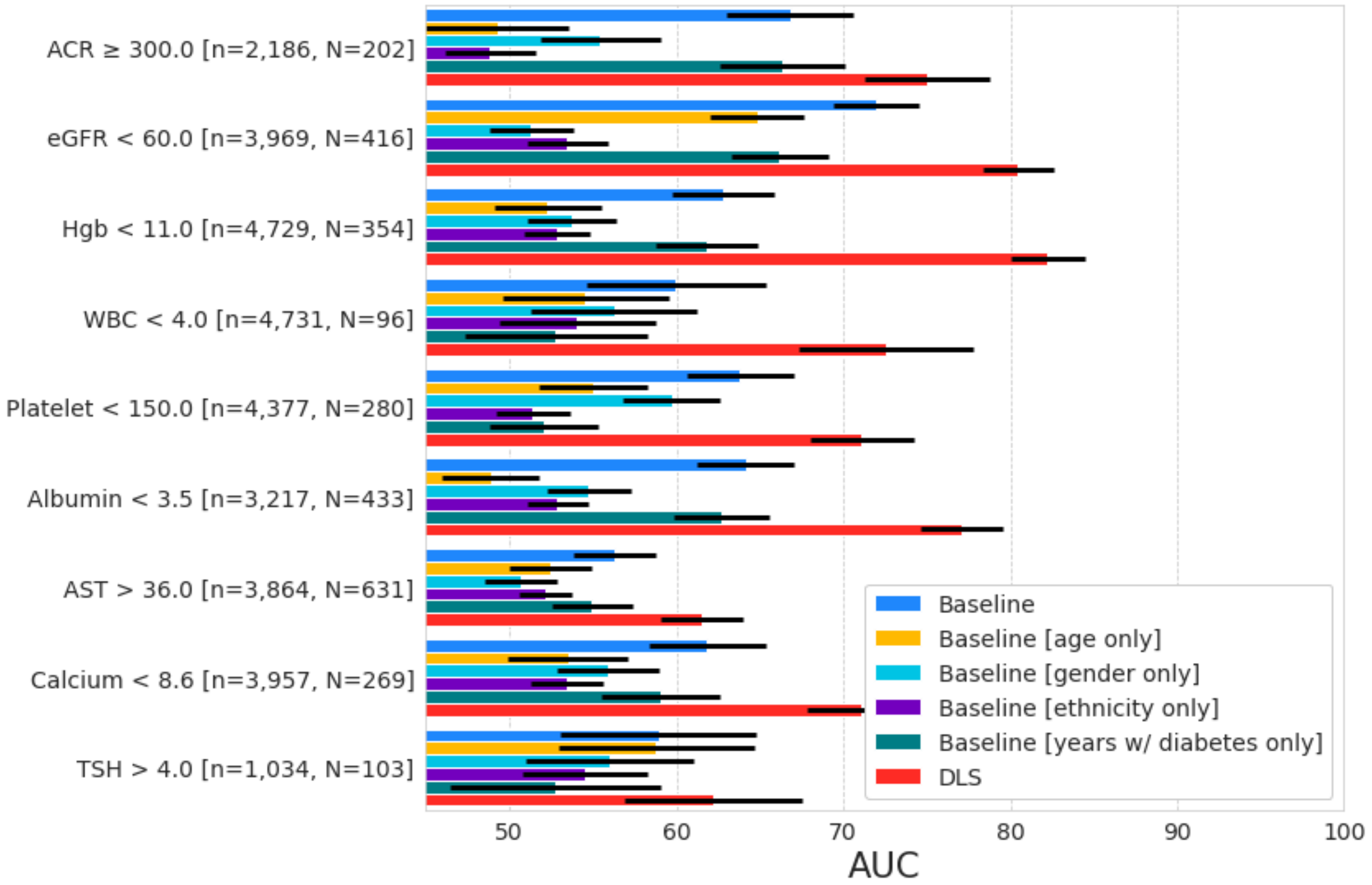

B

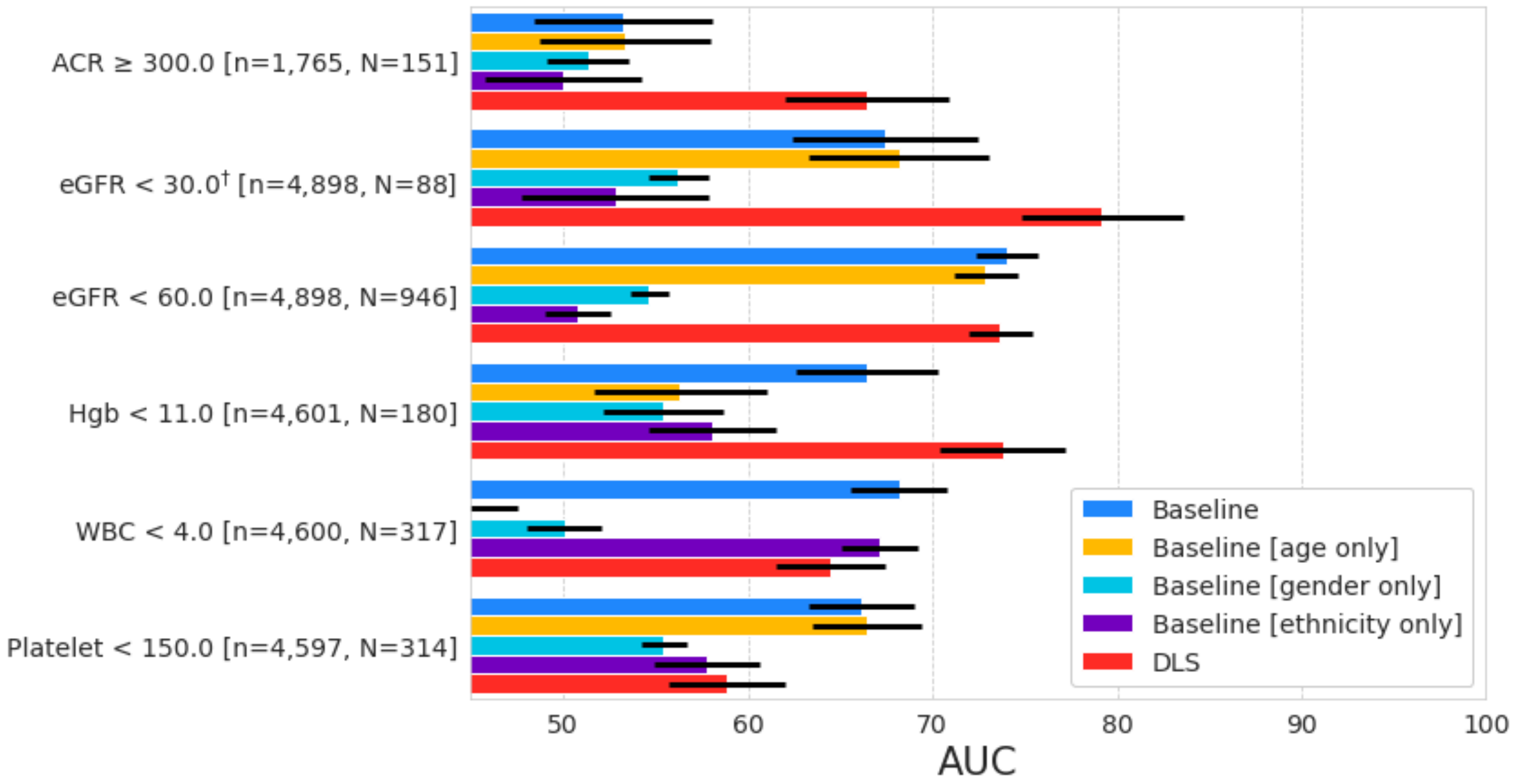

**Supplementary Figure 6. Predictive power of different baseline variables**

Results are presented for (A) validation set A and (B) validation set B. Error bars show 95% confidence intervals computed using DeLong method.

# Supplementary Tables

**Supplementary Table 1. Model versus baseline performance (AUC)**
Results on validation sets A, B and C. †Indicates that the target was prespecified as secondary analysis; all others were prespecified as primary analysis. AUC 95% confidence intervals (CI) computed using the DeLong method. Improvement CI and p-values computed using DeLong paired AUC comparison test.

**A**

| Prediction | n / N (%) | Baseline AUC (CI) | DLS AUC (CI) | Improvement (CI) | p |
|---|---|---|---|---|---|
| ACR ≥ 300.0 | 2186 / 202 (9.2%) | 66.8 (63.0-70.6) | 75.0 (71.2-78.7) | 8.2 (4.3-12.1) | 0.0000 |
| eGFR < 30.0† | 3969 / 107 (2.7%) | 77.0 (72.7-81.3) | 87.7 (84.7-90.7) | 10.7 (6.9-14.6) | 0.0000 |
| eGFR < 60.0 | 3969 / 416 (10.5%) | 72.0 (69.4-74.5) | 80.5 (78.3-82.6) | 8.5 (6.1-10.8) | 0.0000 |
| Hgb < 11.0 | 4729 / 354 (7.5%) | 62.8 (59.8-65.8) | 82.2 (80.0-84.5) | 19.4 (16.4-22.5) | 0.0000 |
| WBC < 4.0 | 4731 / 96 (2.0%) | 60.0 (54.6-65.3) | 72.6 (67.4-77.8) | 12.6 (6.9-18.3) | 0.0000 |
| Platelet < 150.0 | 4377 / 280 (6.4%) | 63.8 (60.6-67.0) | 71.1 (68.0-74.2) | 7.3 (4.3-10.2) | 0.0000 |
| Albumin < 3.5 | 3217 / 433 (13.5%) | 64.1 (61.2-67.0) | 77.1 (74.6-79.5) | 12.9 (9.9-16.0) | 0.0000 |
| AST > 36.0 | 3864 / 631 (16.3%) | 56.3 (53.8-58.8) | 61.5 (59.1-63.9) | 5.2 (2.7-7.8) | 0.0000 |
| Calcium < 8.6 | 3957 / 269 (6.8%) | 61.9 (58.3-65.4) | 71.1 (67.8-74.4) | 9.2 (5.6-12.9) | 0.0000 |
| TSH > 4.0 | 1034 / 103 (10.0%) | 58.9 (53.0-64.8) | 62.2 (56.9-67.5) | 3.3 (-2.8-9.3) | 0.1441 |

**B**

| Prediction | n / N (%) | Baseline AUC (CI) | DLS AUC (CI) | Improvement (CI) | p |
|---|---|---|---|---|---|
| ACR ≥ 300.0 | 1765 / 151 (8.6%) | 53.2 (48.4-58.0) | 66.4 (62.0-70.9) | 13.2 (7.0-19.4) | 0.0000 |
| eGFR < 30.0† | 4898 / 88 (1.8%) | 67.4 (62.4-72.4) | 79.2 (74.8-83.6) | 11.8 (5.4-18.2) | 0.0002 |
| eGFR < 60.0 | 4898 / 946 (19.3%) | 74.0 (72.4-75.7) | 73.7 (72.0-75.4) | -0.4 (-1.9-1.1) | 0.6874 |
| Hgb < 11.0 | 4601 / 180 (3.9%) | 66.5 (62.6-70.3) | 73.8 (70.4-77.2) | 7.3 (2.1-12.6) | 0.0031 |
| WBC < 4.0 | 4600 / 317 (6.9%) | 68.2 (65.6-70.8) | 64.5 (61.5-67.4) | -3.7 (-6.8--0.6) | 0.9899 |
| Platelet < 150.0 | 4597 / 314 (6.8%) | 66.2 (63.3-69.0) | 58.9 (55.7-62.0) | -7.3 (-10.4--4.2) | 1.0000 |

**C**

| Prediction | n / N (%) | Baseline AUC (CI) | DLS AUC (CI) | Improvement (CI) | p |
|---|---|---|---|---|---|
| ACR ≥ 300.0 | 6717 / 541 (8.1%) | 54.4 (51.9-56.9) | 66.1 (63.7-68.5) | 11.7 (8.6-14.9) | 0.0000 |
| eGFR < 30.0† | 9903 / 266 (2.7%) | 59.9 (56.7-63.1) | 76.9 (74.2-79.6) | 17.0 (13.2-20.9) | 0.0000 |
| eGFR < 60.0 | 9903 / 2311 (23.3%) | 70.5 (69.3-71.7) | 71.6 (70.4-72.7) | 1.1 (-0.0-2.1) | 0.0288 |

**Supplementary Table 2. Model versus baseline performance (positive predictive value, PPV)** Fraction of true positives identified in the top 5% of predicted positives in validation sets A, B and C. Model PPV and improvement 95% CI, as well as p-values, computed using bootstrap. †Indicates that the target was prespecified as secondary analysis; all others were prespecified as primary analysis.

**A**

| Prediction | n / N (%) | Baseline PPV (CI) | DLS PPV (CI) | Improvement (CI) | p |
|---|---|---|---|---|---|
| ACR ≥ 300.0 | 2186 / 202 (9.2%) | 22.9 (13.8, 30.3) | 41.3 (31.2, 50.5) | 18.3 (6.4, 29.4) | 0.0020 |
| eGFR < 30.0† | 3969 / 107 (2.7%) | 9.1 (5.6, 13.6) | 19.7 (14.6, 26.3) | 10.6 (5.1, 16.7) | 0.0000 |
| eGFR < 60.0 | 3969 / 416 (10.5%) | 31.8 (26.3, 38.9) | 45.5 (38.4, 52.5) | 13.6 (4.5, 20.7) | 0.0000 |
| Hgb < 11.0 | 4729 / 354 (7.5%) | 13.1 (8.9, 17.4) | 44.1 (37.3, 51.7) | 30.9 (24.6, 39.0) | 0.0000 |
| WBC < 4.0 | 4731 / 96 (2.0%) | 3.4 (1.3, 5.5) | 8.5 (5.5, 12.3) | 5.1 (1.7, 9.7) | 0.0035 |
| Platelet < 150.0 | 4377 / 280 (6.4%) | 8.7 (5.0, 12.4) | 22.0 (16.5, 27.5) | 13.3 (6.9, 19.7) | 0.0000 |
| Albumin < 3.5 | 3217 / 433 (13.5%) | 29.4 (23.1, 37.5) | 56.2 (48.8, 65.6) | 26.9 (16.9, 36.2) | 0.0000 |
| AST > 36.0 | 3864 / 631 (16.3%) | 26.4 (19.7, 33.2) | 36.3 (29.0, 43.0) | 9.8 (2.1, 17.1) | 0.0070 |
| Calcium < 8.6 | 3957 / 269 (6.8%) | 17.8 (12.2, 22.9) | 29.4 (23.4, 36.5) | 11.7 (5.1, 18.8) | 0.0005 |
| TSH > 4.0 | 1034 / 103 (10.0%) | 25.5 (11.8, 35.3) | 17.6 (5.9, 27.5) | -7.8 (-19.6, 9.8) | 0.7800 |

**B**

| Prediction | n / N (%) | Baseline PPV (CI) | DLS PPV (CI) | Improvement (CI) | p |
|---|---|---|---|---|---|
| ACR ≥ 300.0 | 1765 / 151 (8.6%) | 6.8 (3.4, 13.6) | 27.3 (18.2, 37.5) | 20.5 (8.0, 30.7) | 0.0000 |
| eGFR < 30.0† | 4898 / 88 (1.8%) | 7.4 (4.1, 10.7) | 9.4 (6.1, 13.5) | 2.0 (-2.0, 7.0) | 0.1670 |
| eGFR < 60.0 | 4898 / 946 (19.3%) | 54.9 (48.4, 61.1) | 52.0 (45.5, 58.6) | -2.9 (-11.1, 4.5) | 0.7565 |
| Hgb < 11.0 | 4601 / 180 (3.9%) | 3.5 (1.3, 6.5) | 12.2 (8.3, 16.1) | 8.7 (3.5, 13.0) | 0.0000 |
| WBC < 4.0 | 4600 / 317 (6.9%) | 12.2 (7.8, 16.1) | 13.0 (9.6, 18.3) | 0.9 (-3.9, 7.4) | 0.2700 |
| Platelet < 150.0 | 4597 / 314 (6.8%) | 11.4 (7.4, 16.2) | 11.8 (7.9, 16.2) | 0.4 (-5.7, 5.3) | 0.4715 |

**C**

| Prediction | n / N (%) | Baseline PPV (CI) | DLS PPV (CI) | Improvement (CI) | p |
|---|---|---|---|---|---|
| ACR ≥ 300.0 | 6717 / 541 (8.1%) | 12.2 (9.0, 15.5) | 23.9 (18.8, 28.4) | 11.6 (6.0, 16.4) | 0.0000 |
| eGFR < 30.0† | 9903 / 266 (2.7%) | 4.2 (2.6, 6.3) | 11.7 (8.9, 14.5) | 7.5 (4.0, 10.5) | 0.0000 |
| eGFR < 60.0 | 9903 / 2311 (23.3%) | 60.4 (56.2, 64.0) | 59.6 (55.2, 63.8) | -0.8 (-5.3, 3.6) | 0.6415 |

**Supplementary Table 3. Subgroup analysis**

Results for validation sets A, B and C. Rows with N < 25 and predictions for which improvement was below 0 for the full dataset (i.e. “All” subgroup) were omitted. Improvements are highlighted (bold and **) for subgroups where there was a drop of more than 5% between the subgroup and the full dataset; p-values > 0.05 are similarly highlighted. The numerically most meaningful subgroups to examine are those with the “Improvement” column highlighted (large decrease), the “p” column highlighted (lack of statistical significance), and the “N” column indicating a sufficiently large sample size to be meaningful. AUC 95% CI computed using the DeLong method. Improvement CI and p-values computed using DeLong paired AUC comparison test. YrsDM indicates “years with diabetes”.

**A**

| Prediction | Subgroup | n / N (%) | Baseline AUC (CI) | DLS AUC (CI) | Improvement (CI) | p |
|---|---|---|---|---|---|---|
| ACR ≥ 300.0 | All | 2186 / 202 (9.2%) | 66.8 (63.0-70.6) | 75.0 (71.2-78.7) | 8.2 (4.3-12.1) | 0.0000 |
| | Age (0, 50] | 428 / 51 (11.9%) | 65.2 (57.3-73.1) | 68.9 (60.1-77.6) | 3.7 (-3.6-11.0) | **0.1626**** |
| | Age (50, 60] | 655 / 60 (9.2%) | 70.8 (64.2-77.3) | 81.2 (75.6-86.7) | 10.4 (3.5-17.3) | 0.0017 |
| | Age (60, 70] | 534 / 50 (9.4%) | 67.6 (59.8-75.3) | 79.2 (72.6-85.7) | 11.6 (3.4-19.8) | 0.0028 |
| | BMI (0, 25] | 315 / 38 (12.1%) | 67.9 (59.4-76.3) | 79.0 (70.7-87.4) | 11.2 (2.5-19.9) | 0.0059 |
| | BMI (25, 30] | 600 / 59 (9.8%) | 68.0 (60.6-75.4) | 77.4 (70.6-84.1) | 9.3 (2.0-16.7) | 0.0063 |
| | BMI (30, 35] | 527 / 40 (7.6%) | 65.4 (56.3-74.4) | 68.9 (59.8-78.1) | 3.6 (-5.3-12.5) | **0.2138**** |
| | BMI (35, inf) | 525 / 59 (11.2%) | 71.6 (65.1-78.1) | 77.0 (70.3-83.7) | 5.3 (-1.4-12.1) | **0.0616**** |
| | No cataract | 1533 / 145 (9.5%) | 65.9 (61.5-70.4) | 76.5 (72.2-80.7) | 10.5 (6.1-15.0) | 0.0000 |
| | Sex=Female | 821 / 74 (9.0%) | 68.1 (62.4-73.9) | 73.5 (67.3-79.6) | 5.3 (-0.3-10.9) | 0.0310 |
| | Sex=Male | 819 / 95 (11.6%) | 64.3 (58.5-70.1) | 75.9 (70.6-81.2) | 11.6 (4.8-18.5) | 0.0005 |
| | Pupil (0.4, 0.5] | 298 / 43 (14.4%) | 65.2 (56.5-73.9) | 80.3 (74.1-86.5) | 15.1 (5.2-25.1) | 0.0014 |
| | Pupil (0.5, 1.0] | 1410 / 130 (9.2%) | 66.6 (61.9-71.4) | 73.7 (68.8-78.6) | 7.1 (2.8-11.3) | 0.0006 |
| | Race=Hispanic | 1288 / 140 (10.9%) | 69.3 (64.9-73.8) | 78.5 (74.4-82.6) | 9.2 (4.8-13.6) | 0.0000 |
| | YrsDM (0, 5] | 911 / 48 (5.3%) | 64.2 (56.4-72.0) | 69.2 (60.5-77.9) | 5.0 (-3.5-13.5) | **0.1235**** |
| | YrsDM (5, 10] | 454 / 54 (11.9%) | 51.3 (43.2-59.5) | 71.4 (64.7-78.2) | 20.1 (11.3-29.0) | 0.0000 |
| | YrsDM (10, inf) | 493 / 78 (15.8%) | 60.7 (53.9-67.5) | 74.3 (68.6-80.1) | 13.7 (5.8-21.5) | 0.0003 |
| eGFR < 30.0† | All | 3969 / 107 (2.7%) | 77.0 (72.7-81.3) | 87.7 (84.7-90.7) | 10.7 (6.9-14.6) | 0.0000 |
| | Age (50, 60] | 1379 / 39 (2.8%) | 75.6 (67.5-83.7) | 86.4 (79.7-93.1) | 10.8 (4.7-16.8) | 0.0002 |
| | Age (60, 70] | 1139 / 33 (2.9%) | 76.8 (68.8-84.7) | 87.9 (82.6-93.2) | 11.2 (4.3-18.0) | 0.0007 |
| | BMI (0, 25] | 705 / 33 (4.7%) | 75.5 (68.0-83.1) | 84.4 (79.2-89.6) | 8.9 (1.1-16.7) | 0.0128 |
| | BMI (25, 30] | 1281 / 27 (2.1%) | 81.3 (73.6-89.0) | 89.8 (83.8-95.7) | 8.4 (2.1-14.8) | 0.0045 |
| | BMI (30, 35] | 1122 / 25 (2.2%) | 73.4 (62.8-84.0) | 89.2 (83.8-94.6) | 15.8 (6.1-25.5) | 0.0007 |
| | No cataract | 3161 / 83 (2.6%) | 77.1 (72.4-81.8) | 89.0 (85.7-92.4) | 11.9 (7.5-16.4) | 0.0000 |
| | Sex=Female | 1720 / 49 (2.8%) | 81.5 (75.9-87.1) | 89.7 (86.1-93.4) | 8.3 (3.1-13.4) | 0.0008 |
| | Sex=Male | 1700 / 49 (2.9%) | 75.6 (69.0-82.3) | 88.0 (84.0-92.1) | 12.4 (6.7-18.1) | 0.0000 |
| | Pupil (0.4, 0.5] | 623 / 29 (4.7%) | 71.5 (63.9-79.2) | 87.1 (81.9-92.3) | 15.6 (6.7-24.4) | 0.0003 |
| | Pupil (0.5, 1.0] | 2907 / 63 (2.2%) | 78.3 (72.8-83.8) | 88.8 (85.4-92.3) | 10.5 (5.9-15.2) | 0.0000 |
| | Race=Hispanic | 2791 / 71 (2.5%) | 79.3 (74.5-84.1) | 89.6 (86.5-92.6) | 10.2 (5.8-14.7) | 0.0000 |
| | YrsDM (10, inf) | 1051 / 68 (6.5%) | 63.1 (56.7-69.4) | 82.8 (78.8-86.9) | 19.8 (13.0-26.6) | 0.0000 |
| eGFR < 60.0 | All | 3969 / 416 (10.5%) | 72.0 (69.4-74.5) | 80.5 (78.3-82.6) | 8.5 (6.1-10.8) | 0.0000 |
| | Age (0, 50] | 892 / 46 (5.2%) | 71.8 (64.3-79.3) | 86.0 (81.0-91.1) | 14.2 (6.7-21.8) | 0.0001 |
| | Age (50, 60] | 1379 / 129 (9.4%) | 64.2 (58.7-69.7) | 77.9 (73.6-82.3) | 13.8 (8.7-18.8) | 0.0000 |
| | Age (60, 70] | 1139 / 161 (14.1%) | 68.4 (64.2-72.6) | 76.6 (72.5-80.7) | 8.2 (3.5-12.9) | 0.0003 |
| | Age (70, inf) | 266 / 69 (25.9%) | 62.4 (55.2-69.7) | 65.8 (58.1-73.5) | **3.4 (-5.7-12.5)**** | **0.2335**** |
| | BMI (0, 25] | 705 / 97 (13.8%) | 74.3 (69.2-79.4) | 82.8 (78.9-86.8) | 8.5 (3.2-13.9) | 0.0008 |
| | BMI (25, 30] | 1281 / 136 (10.6%) | 74.2 (70.1-78.2) | 82.5 (79.1-85.9) | 8.3 (4.6-12.1) | 0.0000 |
| | BMI (30, 35] | 1122 / 105 (9.4%) | 68.9 (63.5-74.2) | 78.6 (74.3-83.0) | 9.8 (5.0-14.5) | 0.0000 |
| | BMI (35, inf) | 1038 / 105 (10.1%) | 71.9 (66.6-77.1) | 80.7 (76.5-84.9) | 8.8 (4.3-13.4) | 0.0001 |
| | Cataract | 377 / 74 (19.6%) | 65.9 (59.1-72.7) | 75.2 (69.5-80.9) | 9.3 (2.0-16.6) | 0.0065 |
| | No cataract | 3161 / 328 (10.4%) | 71.4 (68.4-74.3) | 80.1 (77.7-82.5) | 8.7 (6.0-11.4) | 0.0000 |
| | Sex=Female | 1720 / 188 (10.9%) | 72.8 (69.1-76.5) | 81.1 (78.0-84.2) | 8.3 (4.9-11.7) | 0.0000 |
| | Sex=Male | 1700 / 197 (11.6%) | 70.7 (66.9-74.4) | 79.3 (76.2-82.5) | 8.7 (5.0-12.3) | 0.0000 |
| | Pupil (0.4, 0.5] | 623 / 99 (15.9%) | 69.1 (63.7-74.5) | 77.9 (72.9-82.8) | 8.8 (2.7-14.9) | 0.0022 |
| | Pupil (0.5, 1.0] | 2907 / 285 (9.8%) | 71.4 (68.2-74.5) | 80.4 (77.9-83.0) | 9.1 (6.3-11.8) | 0.0000 |
| | Race=Black | 234 / 48 (20.5%) | 72.7 (64.7-80.6) | 75.2 (67.4-83.1) | **2.6 (-5.8-11.0)**** | **0.2745**** |
| | Race=Hispanic | 2791 / 271 (9.7%) | 71.4 (68.1-74.6) | 81.0 (78.3-83.6) | 9.6 (6.7-12.6) | 0.0000 |
| | YrsDM (0, 5] | 1844 / 124 (6.7%) | 68.9 (63.9-73.9) | 79.3 (75.5-83.2) | 10.5 (6.2-14.8) | 0.0000 |
| | YrsDM (5, 10] | 915 / 98 (10.7%) | 62.8 (56.9-68.7) | 76.6 (71.7-81.5) | 13.8 (8.1-19.5) | 0.0000 |
| | YrsDM (10, inf) | 1051 / 202 (19.2%) | 65.9 (61.8-69.9) | 77.6 (74.2-80.9) | 11.7 (7.5-15.8) | 0.0000 |
| Hgb < 11.0 | All | 4729 / 354 (7.5%) | 62.8 (59.8-65.8) | 82.2 (80.0-84.5) | 19.4 (16.4-22.5) | 0.0000 |
| | Age (0, 50] | 1162 / 92 (7.9%) | 66.4 (61.2-71.6) | 80.8 (76.0-85.6) | **14.4 (8.9-19.9)**** | 0.0000 |
| | Age (50, 60] | 1692 / 120 (7.1%) | 57.2 (51.8-62.5) | 82.3 (78.4-86.2) | 25.1 (19.6-30.6) | 0.0000 |
| | Age (60, 70] | 1380 / 114 (8.3%) | 64.0 (58.5-69.4) | 82.7 (79.0-86.4) | 18.7 (12.9-24.6) | 0.0000 |
| | Age (70, inf) | 325 / 41 (12.6%) | 63.3 (53.9-72.7) | 80.6 (73.7-87.5) | 17.3 (7.0-27.5) | 0.0005 |
| | BMI (0, 25] | 1072 / 99 (9.2%) | 62.3 (56.6-67.9) | 83.7 (79.7-87.7) | 21.4 (15.2-27.6) | 0.0000 |
| | BMI (25, 30] | 1755 / 125 (7.1%) | 64.9 (59.9-69.8) | 83.8 (80.4-87.3) | 19.0 (13.9-24.0) | 0.0000 |
| | BMI (30, 35] | 1536 / 104 (6.8%) | 62.4 (57.0-67.8) | 81.3 (77.2-85.5) | 18.9 (12.9-25.0) | 0.0000 |
| | BMI (35, inf) | 1446 / 109 (7.5%) | 63.6 (58.2-69.0) | 79.4 (74.9-83.8) | 15.8 (10.2-21.4) | 0.0000 |
| | Cataract | 486 / 58 (11.9%) | 60.1 (52.0-68.3) | 80.3 (74.9-85.7) | 20.1 (10.9-29.4) | 0.0000 |
| | No cataract | 3964 / 285 (7.2%) | 63.3 (60.0-66.5) | 84.3 (82.1-86.6) | 21.1 (17.7-24.4) | 0.0000 |
| | Sex=Female | 2208 / 195 (8.8%) | 59.5 (55.3-63.8) | 79.7 (76.4-82.9) | 20.2 (15.7-24.7) | 0.0000 |
| | Sex=Male | 2111 / 138 (6.5%) | 66.8 (61.9-71.8) | 85.5 (82.3-88.8) | 18.7 (13.4-24.0) | 0.0000 |
| | Pupil (0.4, 0.5] | 719 / 86 (12.0%) | 60.3 (54.0-66.7) | 82.0 (77.7-86.4) | 21.7 (14.7-28.7) | 0.0000 |
| | Pupil (0.5, 1.0] | 3639 / 253 (7.0%) | 64.2 (60.8-67.7) | 81.4 (78.7-84.1) | 17.2 (13.5-20.8) | 0.0000 |
| | Race=Black | 315 / 29 (9.2%) | 71.1 (62.0-80.2) | 76.4 (68.0-84.8) | **5.3 (-4.2-14.8)**** | **0.1361**** |
| | Race=Hispanic | 3505 / 280 (8.0%) | 62.1 (58.8-65.4) | 83.1 (80.6-85.6) | 21.0 (17.4-24.7) | 0.0000 |
| | YrsDM (0, 5] | 2373 / 126 (5.3%) | 57.4 (52.2-62.6) | 82.0 (78.0-86.0) | 24.6 (19.2-30.0) | 0.0000 |
| | YrsDM (5, 10] | 1114 / 85 (7.6%) | 53.8 (47.7-59.9) | 83.1 (79.2-87.0) | 29.3 (22.1-36.5) | 0.0000 |
| | YrsDM (10, inf) | 1283 / 165 (12.9%) | 55.9 (51.5-60.3) | 80.3 (76.9-83.8) | 24.4 (19.3-29.5) | 0.0000 |
| WBC < 4.0 | All | 4731 / 96 (2.0%) | 60.0 (54.6-65.3) | 72.6 (67.4-77.8) | 12.6 (6.9-18.3) | 0.0000 |
| | Age (50, 60] | 1693 / 50 (3.0%) | 61.5 (52.9-70.1) | 73.0 (65.8-80.1) | 11.4 (2.1-20.7) | 0.0080 |
| | Age (60, 70] | 1381 / 34 (2.5%) | 53.6 (44.1-63.1) | 74.1 (64.8-83.4) | 20.5 (9.7-31.2) | 0.0001 |
| | BMI (0, 25] | 1072 / 32 (3.0%) | 57.3 (48.7-66.0) | 78.3 (70.0-86.6) | 21.0 (11.1-30.9) | 0.0000 |
| | BMI (25, 30] | 1756 / 35 (2.0%) | 60.1 (50.8-69.3) | 63.0 (52.6-73.4) | **2.9 (-6.4-12.3)**** | **0.2683**** |
| | BMI (30, 35] | 1538 / 28 (1.8%) | 58.6 (46.7-70.4) | 76.4 (67.0-85.7) | 17.8 (5.0-30.5) | 0.0031 |
| | No cataract | 3966 / 90 (2.3%) | 61.8 (56.2-67.5) | 73.6 (68.3-78.8) | 11.7 (5.8-17.7) | 0.0001 |
| | Sex=Female | 2206 / 42 (1.9%) | 64.9 (57.1-72.8) | 75.5 (68.2-82.7) | 10.5 (2.9-18.2) | 0.0036 |
| | Sex=Male | 2114 / 49 (2.3%) | 54.3 (46.0-62.6) | 69.9 (61.9-77.9) | 15.6 (6.4-24.7) | 0.0004 |
| | Pupil (0.5, 1.0] | 3641 / 83 (2.3%) | 59.3 (53.5-65.1) | 70.0 (64.2-75.8) | 10.7 (4.6-16.8) | 0.0003 |
| | Race=Hispanic | 3507 / 63 (1.8%) | 57.9 (51.7-64.0) | 69.8 (62.4-77.1) | 11.9 (5.3-18.5) | 0.0002 |
| | YrsDM (0, 5] | 2373 / 49 (2.1%) | 61.9 (54.3-69.5) | 70.2 (62.9-77.6) | 8.3 (-0.1-16.8) | 0.0262 |
| | YrsDM (5, 10] | 1115 / 26 (2.3%) | 52.5 (41.8-63.2) | 73.3 (61.7-85.0) | 20.8 (10.5-31.1) | 0.0000 |
| Platelet < 150.0 | All | 4377 / 280 (6.4%) | 63.8 (60.6-67.0) | 71.1 (68.0-74.2) | 7.3 (4.3-10.2) | 0.0000 |
| | Age (0, 50] | 1031 / 45 (4.4%) | 65.1 (56.8-73.4) | 71.7 (62.7-80.7) | 6.6 (-0.1-13.3) | 0.0272 |
| | Age (50, 60] | 1525 / 109 (7.1%) | 61.1 (55.8-66.3) | 74.3 (69.7-79.0) | 13.3 (8.2-18.3) | 0.0000 |
| | Age (60, 70] | 1235 / 88 (7.1%) | 57.1 (50.8-63.4) | 66.3 (60.3-72.4) | 9.2 (2.3-16.2) | 0.0046 |
| | BMI (0, 25] | 845 / 59 (7.0%) | 54.9 (47.2-62.5) | 70.0 (62.7-77.3) | 15.1 (8.4-21.9) | 0.0000 |
| | BMI (25, 30] | 1490 / 88 (5.9%) | 60.0 (54.3-65.6) | 74.0 (68.7-79.3) | 14.0 (9.0-19.1) | 0.0000 |
| | BMI (30, 35] | 1281 / 96 (7.5%) | 67.8 (62.9-72.8) | 72.8 (67.9-77.6) | 4.9 (-0.0-9.9) | 0.0256 |
| | BMI (35, inf) | 1192 / 71 (6.0%) | 69.1 (63.0-75.2) | 72.6 (66.6-78.6) | 3.6 (-2.6-9.7) | **0.1280**** |
| | Cataract | 420 / 37 (8.8%) | 48.4 (39.1-57.7) | 53.8 (43.6-63.9) | 5.4 (-4.5-15.2) | **0.1428**** |
| | No cataract | 3580 / 225 (6.3%) | 64.4 (60.9-67.9) | 71.1 (67.6-74.5) | 6.6 (3.3-10.0) | 0.0001 |
| | Sex=Female | 1959 / 79 (4.0%) | 58.9 (52.7-65.2) | 72.8 (66.8-78.8) | 13.9 (7.3-20.4) | 0.0000 |
| | Sex=Male | 1917 / 169 (8.8%) | 55.7 (51.5-59.9) | 65.7 (61.2-70.1) | 10.0 (5.1-14.9) | 0.0000 |
| | Pupil (0.4, 0.5] | 661 / 41 (6.2%) | 57.1 (48.1-66.2) | 68.4 (59.6-77.3) | 11.3 (4.0-18.6) | 0.0013 |
| | Pupil (0.5, 1.0] | 3261 / 216 (6.6%) | 64.7 (61.1-68.3) | 71.3 (67.7-74.9) | 6.6 (3.1-10.1) | 0.0001 |
| | Race=Hispanic | 3128 / 209 (6.7%) | 61.7 (58.1-65.4) | 71.3 (67.7-74.9) | 9.6 (6.0-13.1) | 0.0000 |
| | YrsDM (0, 5] | 2123 / 130 (6.1%) | 63.6 (58.7-68.5) | 70.6 (65.8-75.4) | 7.0 (2.4-11.6) | 0.0014 |
| | YrsDM (5, 10] | 1000 / 76 (7.6%) | 66.9 (61.5-72.4) | 74.6 (69.3-79.9) | 7.7 (1.6-13.7) | 0.0066 |
| | YrsDM (10, inf) | 1165 / 71 (6.1%) | 59.5 (52.9-66.0) | 66.3 (60.0-72.6) | 6.8 (0.7-13.0) | 0.0146 |
| Albumin < 3.5 | All | 3217 / 433 (13.5%) | 64.1 (61.2-67.0) | 77.1 (74.6-79.5) | 12.9 (9.9-16.0) | 0.0000 |
| | Age (0, 50] | 730 / 105 (14.4%) | 67.1 (61.3-72.8) | 75.5 (70.2-80.8) | 8.5 (2.6-14.3) | 0.0024 |
| | Age (50, 60] | 1111 / 161 (14.5%) | 64.2 (59.3-69.1) | 78.6 (74.5-82.7) | 14.4 (9.6-19.2) | 0.0000 |
| | Age (60, 70] | 829 / 110 (13.3%) | 63.1 (57.5-68.6) | 75.6 (70.8-80.4) | 12.5 (6.1-19.0) | 0.0001 |
| | Age (70, inf) | 176 / 35 (19.9%) | 57.4 (46.3-68.4) | 67.7 (57.7-77.7) | 10.3 (-3.0-23.6) | **0.0640**** |
| | BMI (0, 25] | 562 / 93 (16.5%) | 64.3 (58.1-70.6) | 75.5 (70.1-80.8) | 11.1 (5.1-17.2) | 0.0002 |
| | BMI (25, 30] | 1015 / 147 (14.5%) | 63.4 (58.0-68.7) | 78.4 (74.2-82.6) | 15.0 (9.8-20.2) | 0.0000 |
| | BMI (30, 35] | 843 / 85 (10.1%) | 58.2 (51.3-65.0) | 80.0 (74.7-85.3) | 21.8 (14.6-29.1) | 0.0000 |
| | BMI (35, inf) | 816 / 128 (15.7%) | 62.9 (57.5-68.2) | 76.7 (72.1-81.2) | 13.8 (8.1-19.5) | 0.0000 |
| | Cataract | 288 / 55 (19.1%) | 62.1 (53.2-70.9) | 77.9 (71.2-84.6) | 15.8 (6.8-24.8) | 0.0003 |
| | No cataract | 2485 / 336 (13.5%) | 62.7 (59.3-66.0) | 77.2 (74.3-80.0) | 14.5 (11.1-17.9) | 0.0000 |
| | Sex=Female | 1362 / 170 (12.5%) | 59.9 (55.1-64.6) | 74.3 (70.2-78.4) | 14.5 (9.5-19.4) | 0.0000 |
| | Sex=Male | 1354 / 224 (16.5%) | 63.3 (59.0-67.5) | 77.1 (73.6-80.6) | 13.8 (9.4-18.3) | 0.0000 |
| | Pupil (0.4, 0.5] | 440 / 85 (19.3%) | 63.8 (57.6-70.1) | 76.3 (70.5-82.1) | 12.5 (5.9-19.0) | 0.0001 |
| | Pupil (0.5, 1.0] | 2274 / 288 (12.7%) | 63.1 (59.4-66.7) | 76.8 (73.8-79.8) | 13.7 (9.9-17.6) | 0.0000 |
| | Race=Black | 207 / 33 (15.9%) | 55.6 (44.5-66.8) | 64.9 (54.1-75.7) | 9.3 (-0.4-18.9) | 0.0298 |
| | Race=Hispanic | 2185 / 327 (15.0%) | 62.9 (59.5-66.3) | 78.5 (75.8-81.3) | 15.6 (12.1-19.1) | 0.0000 |
| | YrsDM (0, 5] | 1470 / 154 (10.5%) | 52.6 (47.8-57.3) | 71.3 (66.7-75.8) | 18.7 (12.7-24.7) | 0.0000 |
| | YrsDM (5, 10] | 678 / 95 (14.0%) | 59.3 (53.0-65.6) | 77.0 (72.0-82.0) | 17.6 (10.0-25.3) | 0.0000 |
| | YrsDM (10, inf) | 824 / 178 (21.6%) | 59.8 (55.1-64.5) | 76.5 (72.6-80.4) | 16.7 (11.4-21.9) | 0.0000 |
| AST > 36.0 | All | 3864 / 631 (16.3%) | 56.3 (53.8-58.8) | 61.5 (59.1-63.9) | 5.2 (2.7-7.8) | 0.0000 |
| | Age (0, 50] | 865 / 170 (19.7%) | 59.7 (54.9-64.5) | 67.3 (62.8-71.7) | 7.6 (3.0-12.1) | 0.0006 |
| | Age (50, 60] | 1339 / 218 (16.3%) | 57.2 (53.1-61.3) | 59.0 (55.1-62.9) | 1.8 (-3.2-6.7) | **0.2417**** |
| | Age (60, 70] | 1084 / 178 (16.4%) | 50.7 (46.1-55.4) | 59.2 (54.5-63.9) | 8.5 (2.7-14.2) | 0.0019 |
| | Age (70, inf) | 244 / 32 (13.1%) | 55.0 (43.9-66.1) | 46.6 (34.6-58.6) | **-8.4 (-24.0-7.2)**** | **0.8542**** |
| | BMI (0, 25] | 884 / 142 (16.1%) | 53.5 (48.2-58.7) | 62.7 (57.6-67.8) | 9.2 (3.3-15.2) | 0.0012 |
| | BMI (25, 30] | 1424 / 210 (14.7%) | 55.1 (50.8-59.4) | 61.5 (57.3-65.7) | 6.4 (2.0-10.8) | 0.0023 |
| | BMI (30, 35] | 1214 / 234 (19.3%) | 55.0 (50.8-59.2) | 59.6 (55.4-63.8) | 4.6 (0.5-8.6) | 0.0135 |
| | BMI (35, inf) | 1123 / 213 (19.0%) | 56.8 (52.5-61.1) | 64.0 (60.0-68.1) | 7.2 (3.3-11.2) | 0.0001 |
| | Cataract | 394 / 55 (14.0%) | 53.2 (44.5-61.8) | 53.0 (43.4-62.6) | **-0.2 (-11.1-10.7)**** | **0.5130**** |
| | No cataract | 3067 / 518 (16.9%) | 56.6 (53.8-59.3) | 62.3 (59.7-65.0) | 5.8 (2.9-8.6) | 0.0000 |
| | Sex=Female | 1697 / 281 (16.6%) | 56.3 (52.6-60.0) | 61.4 (57.8-64.9) | 5.1 (1.5-8.7) | 0.0027 |
| | Sex=Male | 1635 / 273 (16.7%) | 56.3 (52.5-60.1) | 63.3 (59.5-67.1) | 7.0 (2.8-11.2) | 0.0005 |
| | Pupil (0.4, 0.5] | 581 / 81 (13.9%) | 53.8 (46.8-60.8) | 60.7 (53.6-67.8) | 6.9 (-1.0-14.9) | 0.0442 |
| | Pupil (0.5, 1.0] | 2830 / 488 (17.2%) | 56.0 (53.2-58.8) | 60.7 (57.9-63.4) | 4.7 (1.7-7.6) | 0.0009 |
| | Race=Hispanic | 2699 / 458 (17.0%) | 56.8 (53.8-59.7) | 63.2 (60.3-66.0) | 6.4 (3.4-9.4) | 0.0000 |
| | YrsDM (0, 5] | 1824 / 342 (18.8%) | 55.6 (52.2-59.0) | 60.6 (57.3-63.9) | 5.0 (1.4-8.5) | 0.0029 |
| | YrsDM (5, 10] | 871 / 140 (16.1%) | 53.1 (48.2-58.1) | 59.0 (53.7-64.2) | 5.8 (-0.3-12.0) | 0.0311 |
| | YrsDM (10, inf) | 1021 / 141 (13.8%) | 46.5 (41.5-51.4) | 59.3 (54.1-64.6) | 12.9 (6.6-19.1) | 0.0000 |
| Calcium < 8.6 | All | 3957 / 269 (6.8%) | 61.9 (58.3-65.4) | 71.1 (67.8-74.4) | 9.2 (5.6-12.9) | 0.0000 |
| | Age (0, 50] | 864 / 66 (7.6%) | 54.2 (46.6-61.8) | 65.1 (57.8-72.4) | 10.9 (3.2-18.6) | 0.0027 |
| | Age (50, 60] | 1360 / 101 (7.4%) | 64.0 (58.0-70.0) | 74.1 (68.5-79.7) | 10.1 (4.3-15.8) | 0.0003 |
| | Age (60, 70] | 1144 / 70 (6.1%) | 62.1 (55.1-69.2) | 75.0 (69.2-80.8) | 12.9 (5.7-20.0) | 0.0002 |
| | BMI (0, 25] | 718 / 56 (7.8%) | 61.4 (53.4-69.5) | 71.3 (64.2-78.4) | 9.8 (1.5-18.2) | 0.0105 |
| | BMI (25, 30] | 1313 / 100 (7.6%) | 66.3 (60.5-72.2) | 76.8 (71.5-82.1) | 10.4 (5.0-15.9) | 0.0001 |
| | BMI (30, 35] | 1115 / 66 (5.9%) | 57.3 (50.0-64.7) | 67.8 (60.9-74.7) | 10.5 (2.6-18.4) | 0.0047 |
| | BMI (35, inf) | 1029 / 65 (6.3%) | 60.5 (53.1-68.0) | 71.1 (64.2-78.0) | 10.6 (3.7-17.5) | 0.0014 |
| | Cataract | 377 / 33 (8.8%) | 66.0 (55.7-76.3) | 75.1 (66.7-83.4) | 9.1 (-1.1-19.3) | 0.0405 |
| | No cataract | 3180 / 220 (6.9%) | 61.5 (57.6-65.5) | 72.2 (68.6-75.9) | 10.7 (6.7-14.7) | 0.0000 |
| | Sex=Female | 1718 / 97 (5.6%) | 57.6 (51.8-63.4) | 67.8 (62.2-73.4) | 10.2 (3.6-16.7) | 0.0012 |
| | Sex=Male | 1701 / 154 (9.1%) | 62.6 (57.8-67.4) | 73.4 (68.9-77.8) | 10.7 (5.7-15.8) | 0.0000 |
| | Pupil (0.4, 0.5] | 634 / 68 (10.7%) | 59.5 (52.5-66.4) | 76.3 (70.1-82.5) | 16.8 (9.3-24.4) | 0.0000 |
| | Pupil (0.5, 1.0] | 2876 / 176 (6.1%) | 60.0 (55.6-64.4) | 67.4 (63.1-71.8) | 7.4 (2.8-12.0) | 0.0008 |
| | Race=Hispanic | 2746 / 217 (7.9%) | 61.7 (57.6-65.7) | 72.5 (68.8-76.2) | 10.9 (6.9-14.9) | 0.0000 |
| | YrsDM (0, 5] | 1848 / 97 (5.2%) | 57.5 (51.9-63.1) | 62.9 (57.0-68.7) | 5.4 (-1.6-12.4) | **0.0638**** |
| | YrsDM (5, 10] | 908 / 63 (6.9%) | 63.8 (57.1-70.5) | 76.7 (71.0-82.5) | 12.9 (5.9-20.0) | 0.0002 |
| | YrsDM (10, inf) | 1081 / 106 (9.8%) | 59.5 (53.3-65.6) | 73.4 (68.0-78.8) | 13.9 (7.4-20.4) | 0.0000 |
| TSH > 4.0 | All | 1034 / 103 (10.0%) | 58.9 (53.0-64.8) | 62.2 (56.9-67.5) | 3.3 (-2.8-9.3) | 0.1441 |
| | Age (50, 60] | 311 / 33 (10.6%) | 56.7 (47.1-66.4) | 61.4 (51.1-71.8) | 4.7 (-5.6-15.0) | **0.1844**** |
| | Age (60, 70] | 258 / 36 (14.0%) | 57.8 (47.5-68.0) | 57.1 (47.9-66.3) | -0.7 (-11.2-9.9) | **0.5500**** |
| | BMI (25, 30] | 345 / 42 (12.2%) | 59.6 (50.6-68.6) | 63.0 (54.4-71.5) | 3.4 (-6.8-13.6) | **0.2587**** |
| | BMI (30, 35] | 286 / 27 (9.4%) | 63.8 (52.0-75.7) | 64.0 (53.5-74.6) | 0.2 (-12.3-12.7) | **0.4893**** |
| | BMI (35, inf) | 276 / 33 (12.0%) | 53.9 (42.6-65.2) | 64.1 (54.7-73.5) | 10.2 (0.2-20.2) | 0.0228 |
| | No cataract | 744 / 71 (9.5%) | 57.0 (50.1-63.9) | 62.0 (55.7-68.2) | 5.0 (-2.3-12.3) | **0.0909**** |
| | Sex=Female | 403 / 53 (13.2%) | 59.4 (51.0-67.7) | 53.9 (46.0-61.7) | **-5.5 (-13.3-2.3)**** | **0.9157**** |
| | Sex=Male | 413 / 36 (8.7%) | 57.5 (47.2-67.7) | 66.4 (57.9-74.8) | 8.9 (-3.8-21.6) | **0.0846**** |
| | Pupil (0.5, 1.0] | 645 / 64 (9.9%) | 59.6 (52.4-66.9) | 61.7 (55.0-68.3) | 2.0 (-4.9-9.0) | **0.2831**** |
| | Race=Hispanic | 611 / 78 (12.8%) | 57.6 (50.9-64.4) | 58.1 (51.6-64.6) | 0.5 (-6.6-7.5) | **0.4500**** |
| | YrsDM (0, 5] | 412 / 43 (10.4%) | 60.4 (51.5-69.2) | 59.6 (50.7-68.5) | -0.8 (-9.9-8.3) | **0.5679**** |
| | YrsDM (10, inf) | 251 / 36 (14.3%) | 62.2 (51.9-72.5) | 59.2 (49.9-68.4) | **-3.0 (-16.2-10.2)**** | **0.6718**** |

**B**

| Prediction | Subgroup | n / N (%) | Baseline AUC (CI) | DLS AUC (CI) | Improvement (CI) | p |
|---|---|---|---|---|---|---|
| ACR ≥ 300.0 | All | 1765 / 151 (8.6%) | 53.2 (48.4-58.0) | 66.4 (62.0-70.9) | 13.2 (7.0-19.4) | 0.0000 |
| | Age (50, 60] | 379 / 25 (6.6%) | 56.0 (45.4-66.6) | 68.2 (56.5-80.0) | 12.2 (-0.4-24.8) | 0.0285 |
| | Age (60, 70] | 562 / 57 (10.1%) | 52.1 (43.8-60.4) | 68.5 (61.8-75.3) | 16.5 (6.0-27.0) | 0.0011 |
| | Age (70, inf) | 606 / 54 (8.9%) | 49.9 (41.6-58.1) | 62.1 (53.6-70.5) | 12.2 (0.3-24.1) | 0.0224 |
| | BMI (0, 25] | 399 / 30 (7.5%) | 49.8 (38.9-60.7) | 63.8 (54.2-73.5) | 14.0 (3.7-24.4) | 0.0040 |
| | BMI (25, 30] | 698 / 59 (8.5%) | 52.5 (45.0-60.0) | 63.1 (56.2-70.1) | 10.6 (1.1-20.1) | 0.0141 |
| | BMI (30, 35] | 738 / 51 (6.9%) | 53.0 (44.0-62.1) | 66.1 (58.8-73.4) | 13.0 (1.2-24.9) | 0.0157 |
| | BMI (35, inf) | 641 / 64 (10.0%) | 47.7 (40.5-54.8) | 67.2 (60.5-74.0) | 19.6 (11.4-27.8) | 0.0000 |
| | Diabetic | 1336 / 142 (10.6%) | 50.5 (45.5-55.5) | 64.1 (59.4-68.9) | 13.6 (7.0-20.3) | 0.0000 |
| | No IOL | 1681 / 146 (8.7%) | 54.0 (49.2-58.8) | 67.2 (62.7-71.7) | 13.2 (6.8-19.5) | 0.0000 |
| | Sex=Male | 1571 / 140 (8.9%) | 52.7 (47.8-57.6) | 67.3 (62.8-71.8) | 14.6 (7.7-21.5) | 0.0000 |
| | Pupil (0.4, 0.5] | 574 / 55 (9.6%) | 57.4 (49.2-65.7) | 64.8 (57.1-72.4) | **7.3 (-4.1-18.7)**** | **0.1037**** |
| | Pupil (0.5, 1.0] | 1030 / 77 (7.5%) | 50.5 (44.0-56.9) | 66.3 (60.0-72.5) | 15.8 (8.1-23.4) | 0.0000 |
| | Race=Black | 911 / 78 (8.6%) | 55.2 (48.9-61.4) | 65.9 (59.5-72.3) | 10.7 (2.5-19.0) | 0.0053 |
| | Race=White | 803 / 69 (8.6%) | 52.7 (45.6-59.7) | 68.8 (62.3-75.3) | 16.1 (9.2-23.0) | 0.0000 |
| eGFR < 30.0† | All | 4898 / 88 (1.8%) | 67.4 (62.4-72.4) | 79.2 (74.8-83.6) | 11.8 (5.4-18.2) | 0.0002 |
| | Age (60, 70] | 1439 / 26 (1.8%) | 61.0 (51.0-71.0) | 77.2 (69.4-85.0) | 16.2 (5.0-27.4) | 0.0023 |
| | Age (70, inf) | 1474 / 50 (3.4%) | 61.0 (52.0-69.9) | 73.2 (66.5-79.9) | 12.2 (1.6-22.8) | 0.0120 |
| | BMI (0, 25] | 1349 / 25 (1.9%) | 74.5 (66.4-82.6) | 85.2 (78.2-92.2) | 10.7 (-0.5-21.9) | 0.0301 |
| | BMI (25, 30] | 2145 / 36 (1.7%) | 63.3 (54.9-71.8) | 77.7 (70.9-84.6) | 14.4 (5.2-23.6) | 0.0011 |
| | BMI (30, 35] | 1962 / 26 (1.3%) | 73.8 (65.9-81.7) | 80.4 (72.3-88.5) | **6.6 (-6.2-19.4)**** | **0.1568**** |
| | Diabetic | 2132 / 69 (3.2%) | 62.0 (56.0-68.1) | 75.0 (69.5-80.5) | 13.0 (5.6-20.4) | 0.0003 |
| | No IOL | 4718 / 84 (1.8%) | 68.2 (63.2-73.2) | 79.5 (75.0-84.0) | 11.3 (4.7-17.9) | 0.0004 |
| | Sex=Male | 4169 / 86 (2.1%) | 63.0 (57.2-68.7) | 77.8 (73.2-82.5) | 14.9 (7.6-22.1) | 0.0000 |
| | Pupil (0.4, 0.5] | 1384 / 38 (2.7%) | 67.0 (58.6-75.5) | 77.2 (71.2-83.2) | 10.2 (0.2-20.1) | 0.0225 |
| | Pupil (0.5, 1.0] | 3154 / 41 (1.3%) | 66.2 (59.3-73.0) | 81.5 (74.9-88.1) | 15.3 (6.4-24.3) | 0.0004 |
| | Race=Black | 2682 / 54 (2.0%) | 70.4 (63.9-76.9) | 76.9 (70.7-83.2) | **6.6 (-2.5-15.6)**** | **0.0778**** |
| | Race=White | 2129 / 34 (1.6%) | 68.7 (61.0-76.4) | 81.7 (75.7-87.8) | 13.0 (5.8-20.3) | 0.0002 |
| Hgb < 11.0 | All | 4601 / 180 (3.9%) | 66.5 (62.6-70.3) | 73.8 (70.4-77.2) | 7.3 (2.1-12.6) | 0.0031 |
| | Age (0, 50] | 761 / 33 (4.3%) | 80.4 (75.1-85.8) | 78.3 (72.0-84.7) | **-2.1 (-10.0-5.9)**** | **0.6960**** |
| | Age (50, 60] | 1091 / 29 (2.7%) | 72.1 (64.4-79.8) | 69.3 (59.0-79.6) | **-2.8 (-15.8-10.2)**** | **0.6624**** |
| | Age (60, 70] | 1359 / 46 (3.4%) | 58.7 (51.0-66.5) | 69.7 (62.1-77.3) | 11.0 (-0.4-22.3) | 0.0293 |
| | Age (70, inf) | 1386 / 73 (5.3%) | 64.4 (57.0-71.8) | 74.6 (69.9-79.3) | 10.2 (1.2-19.1) | 0.0129 |
| | BMI (0, 25] | 1301 / 59 (4.5%) | 68.8 (62.5-75.2) | 74.5 (69.1-80.0) | 5.7 (-3.0-14.4) | **0.0994**** |
| | BMI (25, 30] | 2034 / 69 (3.4%) | 68.8 (63.0-74.5) | 73.3 (67.9-78.8) | 4.5 (-3.6-12.6) | **0.1357**** |
| | BMI (30, 35] | 1844 / 48 (2.6%) | 65.3 (57.3-73.3) | 69.7 (62.8-76.6) | 4.4 (-6.9-15.7) | **0.2222**** |
| | BMI (35, inf) | 1517 / 45 (3.0%) | 67.3 (59.8-74.8) | 71.4 (64.7-78.2) | 4.1 (-5.9-14.2) | **0.2101**** |
| | Diabetic | 1962 / 105 (5.4%) | 61.1 (56.0-66.1) | 70.5 (65.5-75.4) | 9.4 (2.1-16.7) | 0.0058 |
| | Non-diabetic | 2616 / 75 (2.9%) | 74.4 (69.0-79.9) | 76.4 (71.7-81.1) | **2.0 (-5.2-9.1)**** | **0.2957**** |
| | No IOL | 4426 / 173 (3.9%) | 66.7 (62.8-70.6) | 74.3 (70.8-77.7) | 7.6 (2.3-12.8) | 0.0025 |
| | Sex=Female | 666 / 44 (6.6%) | 45.4 (38.1-52.8) | 67.4 (59.6-75.2) | 22.0 (11.6-32.4) | 0.0000 |
| | Sex=Male | 3905 / 134 (3.4%) | 66.4 (61.7-71.2) | 74.9 (70.9-78.8) | 8.4 (2.0-14.9) | 0.0052 |
| | Pupil (0.4, 0.5] | 1312 / 60 (4.6%) | 67.5 (61.2-73.9) | 71.9 (66.0-77.8) | 4.4 (-5.0-13.7) | **0.1788**** |
| | Pupil (0.5, 1.0] | 2943 / 100 (3.4%) | 64.7 (59.3-70.1) | 75.1 (70.5-79.7) | 10.4 (3.4-17.4) | 0.0019 |
| | Race=Black | 2467 / 123 (5.0%) | 65.2 (61.0-69.5) | 69.7 (65.2-74.2) | 4.5 (-1.2-10.3) | **0.0616**** |
| | Race=White | 2044 / 52 (2.5%) | 65.4 (58.6-72.2) | 80.3 (75.0-85.6) | 15.0 (6.4-23.5) | 0.0003 |

C

| Prediction | Subgroup | n / N (%) | Baseline AUC (CI) | DLS AUC (CI) | Improvement (CI) | p |
|---|---|---|---|---|---|---|
| ACR ≥ 300.0 | All | 6717 / 541 (8.1%) | 54.4 (51.9-56.9) | 66.1 (63.7-68.5) | 11.7 (8.6-14.9) | 0.0000 |
| | Age (0, 50] | 738 / 38 (5.1%) | 55.7 (46.2-65.2) | 74.3 (66.1-82.4) | 18.5 (7.0-30.1) | 0.0008 |
| | Age (50, 60] | 1506 / 131 (8.7%) | 53.9 (48.6-59.2) | 68.6 (63.8-73.3) | 14.7 (7.9-21.4) | 0.0000 |
| | Age (60, 70] | 2957 / 226 (7.6%) | 49.3 (45.4-53.1) | 65.1 (61.3-68.8) | 15.8 (10.9-20.7) | 0.0000 |
| | Age (70, inf) | 1238 / 128 (10.3%) | 52.1 (46.6-57.6) | 63.7 (58.7-68.7) | 11.6 (4.9-18.2) | 0.0003 |
| | Sex=Male | 5891 / 483 (8.2%) | 53.7 (51.0-56.4) | 66.9 (64.4-69.4) | 13.3 (9.9-16.6) | 0.0000 |
| | Pupil (0.0, 0.4] | 324 / 37 (11.4%) | 57.4 (47.8-67.0) | 62.2 (51.8-72.5) | **4.8 (-8.5-18.0)**** | **0.2417**** |
| | Pupil (0.4, 0.5] | 1349 / 133 (9.9%) | 53.8 (48.6-58.9) | 66.1 (61.1-71.0) | 12.3 (5.1-19.5) | 0.0004 |
| | Pupil (0.5, 1.0] | 4877 / 366 (7.5%) | 53.4 (50.3-56.5) | 65.4 (62.5-68.3) | 12.0 (8.2-15.7) | 0.0000 |
| | Race=Black | 1347 / 82 (6.1%) | 57.6 (51.4-63.8) | 64.6 (58.3-71.0) | 7.1 (-1.8-15.9) | **0.0588**** |
| | Race=White | 1344 / 81 (6.0%) | 54.6 (48.5-60.7) | 62.5 (56.8-68.3) | 7.9 (1.4-14.4) | 0.0083 |
| eGFR < 30.0† | All | 9903 / 266 (2.7%) | 59.9 (56.7-63.1) | 76.9 (74.2-79.6) | 17.0 (13.2-20.9) | 0.0000 |
| | Age (50, 60] | 2326 / 41 (1.8%) | 61.2 (52.5-69.9) | 77.7 (70.9-84.5) | 16.5 (6.9-26.1) | 0.0004 |
| | Age (60, 70] | 4773 / 115 (2.4%) | 50.0 (44.6-55.4) | 76.1 (71.7-80.5) | 26.1 (19.7-32.5) | 0.0000 |
| | Age (70, inf) | 2063 / 107 (5.2%) | 46.1 (40.3-51.9) | 69.8 (65.1-74.5) | 23.7 (15.6-31.9) | 0.0000 |
| | Sex=Male | 9363 / 254 (2.7%) | 60.6 (57.3-63.9) | 77.2 (74.5-79.9) | 16.6 (12.7-20.5) | 0.0000 |
| | Pupil (0.4, 0.5] | 2185 / 68 (3.1%) | 57.2 (51.0-63.3) | 76.2 (71.3-81.2) | 19.1 (11.2-27.0) | 0.0000 |
| | Pupil (0.5, 1.0] | 7705 / 183 (2.4%) | 59.8 (55.9-63.8) | 77.6 (74.3-80.8) | 17.7 (13.2-22.3) | 0.0000 |
| | Race=Black | 1998 / 47 (2.4%) | 68.5 (61.9-75.2) | 71.0 (63.6-78.3) | **2.4 (-7.9-12.8)**** | **0.3209**** |
| eGFR < 60.0 | All | 9903 / 2311 (23.3%) | 70.5 (69.3-71.7) | 71.6 (70.4-72.7) | 1.1 (-0.0-2.1) | 0.0288 |
| | Age (0, 50] | 1063 / 61 (5.7%) | 52.7 (45.5-59.9) | 68.1 (61.0-75.3) | 15.4 (7.8-23.0) | 0.0000 |
| | Age (50, 60] | 2326 / 279 (12.0%) | 55.1 (51.6-58.7) | 65.0 (61.6-68.4) | 9.9 (5.6-14.1) | 0.0000 |
| | Age (60, 70] | 4773 / 1100 (23.0%) | 55.0 (53.0-57.0) | 60.8 (58.9-62.7) | 5.8 (3.3-8.2) | 0.0000 |
| | Age (70, inf) | 2063 / 940 (45.6%) | 61.4 (59.0-63.8) | 62.3 (59.9-64.7) | 0.9 (-1.5-3.3) | **0.2261**** |
| | Sex=Female | 454 / 77 (17.0%) | 76.4 (70.1-82.8) | 68.9 (61.8-76.0) | **-7.5 (-13.2--1.9)**** | **0.9958**** |
| | Sex=Male | 9363 / 2227 (23.8%) | 70.7 (69.5-71.9) | 71.4 (70.2-72.6) | 0.7 (-0.4-1.8) | **0.1165**** |
| | Pupil (0.0, 0.4] | 475 / 143 (30.1%) | 71.4 (66.4-76.3) | 68.5 (63.3-73.7) | -2.8 (-7.7-2.1) | **0.8715**** |
| | Pupil (0.4, 0.5] | 2185 / 615 (28.1%) | 69.8 (67.3-72.2) | 69.4 (67.0-71.9) | -0.4 (-2.7-2.0) | **0.6178**** |
| | Pupil (0.5, 1.0] | 7705 / 1682 (21.8%) | 70.3 (68.9-71.7) | 71.2 (69.9-72.6) | 0.9 (-0.3-2.2) | **0.0685**** |
| | Race=Black | 1998 / 414 (20.7%) | 70.4 (67.6-73.1) | 68.0 (65.2-70.9) | -2.3 (-5.1-0.4) | **0.9512**** |
| | Race=White | 2200 / 453 (20.6%) | 73.0 (70.4-75.5) | 72.2 (69.6-74.8) | -0.8 (-3.0-1.5) | **0.7500**** |

**Supplementary Table 4. Lab/vital date delta sensitivity analysis**

Results for validation set A. Improvements are highlighted (bold and **) for subgroups where there was a drop of more than 5% between the subgroup and the full dataset (i.e. “All” subgroup); p-values > 0.05 are similarly highlighted. AUC 95% CI computed using the DeLong method. Improvement CI and p-values computed using DeLong paired AUC comparison test.

| Prediction | Subgroup | n / N (%) | Baseline AUC (CI) | DLS AUC (CI) | Improvement (CI) | p |
|---|---|---|---|---|---|---|
| ACR ≥ 300.0 | All | 2186 / 202 (9.2%) | 66.8 (63.0-70.6) | 75.0 (71.2-78.7) | 8.2 (4.3-12.1) | 0.0000 |
| | Time delta < 90 | 1145 / 116 (10.1%) | 67.9 (63.0-72.8) | 76.2 (71.4-81.0) | 8.3 (3.5-13.1) | 0.0003 |
| | Time delta < 30 | 660 / 65 (9.8%) | 66.7 (60.1-73.3) | 72.5 (65.9-79.1) | 5.8 (-1.2-12.8) | **0.0525**** |
| eGFR < 60.0 | All | 3969 / 416 (10.5%) | 72.0 (69.4-74.5) | 80.5 (78.3-82.6) | 8.5 (6.1-10.8) | 0.0000 |
| | Time delta < 90 | 2661 / 315 (11.8%) | 71.5 (68.5-74.6) | 80.2 (77.8-82.7) | 8.7 (6.0-11.4) | 0.0000 |
| | Time delta < 30 | 1522 / 190 (12.5%) | 70.1 (66.0-74.2) | 81.0 (77.9-84.2) | 11.0 (7.5-14.5) | 0.0000 |
| Hgb < 11.0 | All | 4729 / 354 (7.5%) | 62.8 (59.8-65.8) | 82.2 (80.0-84.5) | 19.4 (16.4-22.5) | 0.0000 |
| | Time delta < 90 | 3463 / 304 (8.8%) | 65.0 (61.9-68.2) | 83.9 (81.7-86.1) | 18.8 (15.5-22.1) | 0.0000 |
| | Time delta < 30 | 2111 / 212 (10.0%) | 62.9 (59.1-66.7) | 83.3 (80.7-86.0) | 20.5 (16.3-24.6) | 0.0000 |
| WBC < 4.0 | All | 4731 / 96 (2.0%) | 60.0 (54.6-65.3) | 72.6 (67.4-77.8) | 12.6 (6.9-18.3) | 0.0000 |
| | Time delta < 90 | 3464 / 83 (2.4%) | 61.7 (55.6-67.8) | 74.0 (68.4-79.6) | 12.3 (6.2-18.4) | 0.0000 |
| | Time delta < 30 | 2111 / 58 (2.7%) | 60.2 (53.6-66.8) | 74.6 (68.6-80.6) | 14.4 (6.6-22.2) | 0.0002 |
| Platelet < 150.0 | All | 4377 / 280 (6.4%) | 63.8 (60.6-67.0) | 71.1 (68.0-74.2) | 7.3 (4.3-10.2) | 0.0000 |
| | Time delta < 90 | 3120 / 203 (6.5%) | 64.4 (60.6-68.1) | 70.5 (66.9-74.1) | 6.1 (2.4-9.8) | 0.0006 |
| | Time delta < 30 | 1912 / 123 (6.4%) | 64.4 (59.7-69.1) | 72.3 (67.8-76.7) | 7.9 (3.0-12.7) | 0.0007 |
| Albumin < 3.5 | All | 3217 / 433 (13.5%) | 64.1 (61.2-67.0) | 77.1 (74.6-79.5) | 12.9 (9.9-16.0) | 0.0000 |
| | Time delta < 90 | 1953 / 276 (14.1%) | 66.4 (62.7-70.0) | 81.1 (78.2-83.9) | 14.7 (11.0-18.4) | 0.0000 |
| | Time delta < 30 | 1057 / 137 (13.0%) | 63.5 (58.2-68.8) | 79.3 (75.1-83.5) | 15.8 (10.3-21.3) | 0.0000 |
| AST > 36.0 | All | 3864 / 631 (16.3%) | 56.3 (53.8-58.8) | 61.5 (59.1-63.9) | 5.2 (2.7-7.8) | 0.0000 |
| | Time delta < 90 | 2511 / 422 (16.8%) | 56.3 (53.2-59.4) | 63.1 (60.2-66.1) | 6.8 (3.5-10.1) | 0.0000 |
| | Time delta < 30 | 1345 / 228 (17.0%) | 58.1 (54.0-62.2) | 64.3 (60.3-68.2) | 6.2 (1.7-10.6) | 0.0032 |
| Calcium < 8.6 | All | 3957 / 269 (6.8%) | 61.9 (58.3-65.4) | 71.1 (67.8-74.4) | 9.2 (5.6-12.9) | 0.0000 |
| | Time delta < 90 | 2663 / 189 (7.1%) | 64.2 (60.0-68.4) | 73.4 (69.6-77.2) | 9.2 (5.0-13.4) | 0.0000 |
| | Time delta < 30 | 1532 / 95 (6.2%) | 63.7 (58.0-69.5) | 75.0 (69.7-80.3) | 11.3 (5.3-17.2) | 0.0001 |
| TSH > 4.0 | All | 1034 / 103 (10.0%) | 58.9 (53.0-64.8) | 62.2 (56.9-67.5) | 3.3 (-2.8-9.3) | 0.1441 |
| | Time delta < 90 | 517 / 60 (11.6%) | 62.6 (54.9-70.3) | 64.6 (58.0-71.3) | 2.0 (-6.0-10.1) | **0.3106**** |
| | Time delta < 30 | 250 / 31 (12.4%) | 64.7 (54.9-74.4) | 62.0 (52.1-71.8) | **-2.7 (-14.7-9.3)**** | **0.6727**** |

**Supplementary Table 5. Model versus baseline on complete set of prediction targets**
Results for validation sets A, B and C. Table D lists units for all measurements. AUC 95% CI computed using the DeLong method. Improvement CI and p-values computed using DeLong paired AUC comparison test.

**A**

| Prediction | n / N (%) | Baseline AUC (CI) | DLS AUC (CI) | Improvement (CI) | p |
|---|---|---|---|---|---|
| ACR ≥ 30.0 | 2186 / 761 (34.8%) | 63.2 (60.7-65.6) | 71.0 (68.7-73.3) | 7.8 (5.3-10.4) | 0.0000 |
| ACR ≥ 300.0 | 2186 / 202 (9.2%) | 66.8 (63.0-70.6) | 75.0 (71.2-78.7) | 8.2 (4.3-12.1) | 0.0000 |
| ACR ≥ 1500.0 | 2186 / 61 (2.8%) | 70.9 (64.7-77.1) | 83.7 (77.8-89.6) | 12.8 (6.9-18.6) | 0.0000 |
| Albumin < 3.5 | 3217 / 433 (13.5%) | 64.1 (61.2-67.0) | 77.1 (74.6-79.5) | 12.9 (9.9-16.0) | 0.0000 |
| ALT > 29.0 | 3934 / 1445 (36.7%) | 61.6 (59.8-63.5) | 66.8 (65.0-68.5) | 5.1 (3.6-6.6) | 0.0000 |
| AST > 36.0 | 3864 / 631 (16.3%) | 56.3 (53.8-58.8) | 61.5 (59.1-63.9) | 5.2 (2.7-7.8) | 0.0000 |
| BMI ≥ 25.0 | 5913 / 5190 (87.8%) | 63.7 (61.5-65.9) | 81.9 (80.4-83.5) | 18.2 (16.0-20.5) | 0.0000 |
| BMI ≥ 30.0 | 5913 / 3307 (55.9%) | 63.2 (61.7-64.6) | 78.5 (77.4-79.7) | 15.4 (13.9-16.8) | 0.0000 |
| BMI ≥ 35.0 | 5913 / 1630 (27.6%) | 63.9 (62.4-65.4) | 78.3 (77.1-79.6) | 14.4 (12.8-16.0) | 0.0000 |
| BMI ≥ 40.0 | 5913 / 733 (12.4%) | 65.7 (63.7-67.7) | 80.1 (78.5-81.8) | 14.4 (12.4-16.4) | 0.0000 |
| BUN > 20.0 | 4561 / 870 (19.1%) | 68.8 (67.0-70.7) | 75.4 (73.7-77.2) | 6.6 (4.8-8.4) | 0.0000 |
| Calcium < 8.6 | 3957 / 269 (6.8%) | 61.9 (58.3-65.4) | 71.1 (67.8-74.4) | 9.2 (5.6-12.9) | 0.0000 |
| Creatinine > 1.2 | 3969 / 435 (11.0%) | 73.1 (70.7-75.6) | 79.9 (77.8-82.0) | 6.8 (4.6-8.9) | 0.0000 |
| Diastolic BP ≥ 80.0 | 5025 / 943 (18.8%) | 65.8 (63.9-67.7) | 67.8 (66.0-69.7) | 2.0 (0.6-3.4) | 0.0021 |
| Diastolic BP ≥ 90.0 | 5025 / 175 (3.5%) | 66.3 (62.1-70.5) | 71.4 (67.7-75.2) | 5.1 (1.9-8.4) | 0.0011 |
| eGFR < 15.0 | 3969 / 56 (1.4%) | 74.4 (68.0-80.8) | 87.5 (83.3-91.7) | 13.1 (7.2-18.9) | 0.0000 |
| eGFR < 30.0 | 3969 / 107 (2.7%) | 77.0 (72.7-81.3) | 87.7 (84.7-90.7) | 10.7 (6.9-14.6) | 0.0000 |
| eGFR < 60.0 | 3969 / 416 (10.5%) | 72.0 (69.4-74.5) | 80.5 (78.3-82.6) | 8.5 (6.1-10.8) | 0.0000 |
| eGFR < 90.0 | 3969 / 1257 (31.7%) | 73.3 (71.6-75.0) | 78.0 (76.5-79.5) | 4.7 (3.5-5.9) | 0.0000 |
| HbA1c ≥ 6.5 | 3273 / 2680 (81.9%) | 68.7 (66.4-71.0) | 69.1 (66.9-71.4) | 0.5 (-2.2-3.1) | 0.3678 |
| HbA1c ≥ 7.0 | 3273 / 2143 (65.5%) | 71.6 (69.7-73.5) | 71.9 (70.1-73.7) | 0.3 (-1.7-2.3) | 0.3789 |
| HbA1c ≥ 8.0 | 3273 / 1457 (44.5%) | 70.2 (68.5-72.0) | 73.4 (71.7-75.1) | 3.2 (1.3-5.0) | 0.0003 |
| HbA1c ≥ 9.0 | 3273 / 968 (29.6%) | 67.2 (65.3-69.2) | 72.6 (70.7-74.5) | 5.4 (3.5-7.2) | 0.0000 |
| HCT < 39.0 | 4741 / 1722 (36.3%) | 67.5 (65.9-69.0) | 76.1 (74.7-77.5) | 8.6 (7.2-10.0) | 0.0000 |
| HDL ≥ 45.0 | 3365 / 1478 (43.9%) | 64.3 (62.5-66.2) | 68.0 (66.2-69.8) | 3.7 (2.1-5.2) | 0.0000 |
| HDL ≥ 60.0 | 3365 / 381 (11.3%) | 64.1 (61.1-67.1) | 69.8 (67.1-72.4) | 5.7 (2.9-8.5) | 0.0000 |
| Hgb < 11.0 | 4729 / 354 (7.5%) | 62.8 (59.8-65.8) | 82.2 (80.0-84.5) | 19.4 (16.4-22.5) | 0.0000 |
| Hgb < 12.5 | 4729 / 1138 (24.1%) | 66.4 (64.6-68.1) | 78.2 (76.7-79.7) | 11.8 (10.1-13.6) | 0.0000 |
| INR < 1.1 | 276 / 129 (46.7%) | 64.7 (58.3-71.1) | 66.8 (60.4-73.2) | 2.1 (-3.9-8.1) | 0.2454 |
| LDL ≥ 100.0 | 3373 / 1620 (48.0%) | 60.9 (59.0-62.8) | 60.2 (58.3-62.1) | -0.8 (-2.5-1.0) | 0.8071 |
| LDL ≥ 130.0 | 3373 / 746 (22.1%) | 58.0 (55.7-60.3) | 59.2 (56.9-61.5) | 1.2 (-1.2-3.6) | 0.1696 |
| LDL ≥ 160.0 | 3373 / 255 (7.6%) | 56.1 (52.4-59.7) | 61.2 (57.5-64.9) | 5.2 (1.3-9.0) | 0.0040 |
| LDL ≥ 190.0 | 3373 / 84 (2.5%) | 57.9 (51.6-64.1) | 62.0 (55.8-68.2) | 4.1 (-2.7-11.0) | 0.1192 |
| Mean arterial pressure ≥ 80.0 | 4672 / 4093 (87.6%) | 59.6 (57.2-62.0) | 63.2 (60.8-65.6) | 3.6 (1.2-6.0) | 0.0017 |
| Mean arterial pressure ≥ 90.0 | 4672 / 2453 (52.5%) | 59.2 (57.6-60.9) | 63.0 (61.4-64.6) | 3.8 (2.4-5.2) | 0.0000 |
| Mean arterial pressure ≥ 110.0 | 4672 / 131 (2.8%) | 64.7 (59.9-69.5) | 69.4 (64.8-74.0) | 4.7 (0.2-9.2) | 0.0194 |
| non-HDL ≥ 130.0 | 2939 / 1447 (49.2%) | 60.4 (58.4-62.4) | 60.3 (58.3-62.3) | -0.1 (-1.9-1.6) | 0.5578 |
| non-HDL ≥ 160.0 | 2939 / 753 (25.6%) | 58.6 (56.3-61.0) | 59.4 (57.1-61.8) | 0.8 (-1.4-3.0) | 0.2486 |
| Platelet < 100.0 | 4377 / 64 (1.5%) | 69.4 (63.9-74.8) | 84.9 (80.4-89.4) | 15.5 (9.6-21.4) | 0.0000 |
| Platelet < 150.0 | 4377 / 280 (6.4%) | 63.8 (60.6-67.0) | 71.1 (68.0-74.2) | 7.3 (4.3-10.2) | 0.0000 |
| Potassium < 3.5 | 3990 / 89 (2.2%) | 55.0 (48.3-61.7) | 54.1 (47.2-60.9) | -0.9 (-8.0-6.2) | 0.6004 |
| Potassium > 5.0 | 3990 / 176 (4.4%) | 67.1 (63.0-71.1) | 73.2 (69.4-77.0) | 6.1 (2.4-9.8) | 0.0006 |
| Pulse pressure ≥ 40.0 | 4672 / 4359 (93.3%) | 64.6 (61.6-67.6) | 66.7 (63.5-69.9) | 2.1 (-0.2-4.4) | 0.0377 |
| Pulse pressure ≥ 55.0 | 4672 / 2825 (60.5%) | 67.5 (65.9-69.1) | 68.3 (66.7-69.8) | 0.8 (-0.2-1.8) | 0.0548 |
| Pulse pressure ≥ 65.0 | 4672 / 1464 (31.3%) | 69.2 (67.6-70.8) | 70.2 (68.6-71.7) | 0.9 (-0.1-2.0) | 0.0376 |
| RDW > 14.5 | 4734 / 1223 (25.8%) | 59.7 (57.9-61.6) | 66.9 (65.2-68.7) | 7.2 (5.4-9.0) | 0.0000 |
| Sodium < 136.0 | 3959 / 1047 (26.4%) | 60.7 (58.7-62.7) | 64.9 (63.0-66.8) | 4.2 (2.4-6.0) | 0.0000 |
| Systolic BP ≥ 120.0 | 5010 / 3807 (76.0%) | 60.5 (58.7-62.4) | 63.2 (61.3-65.1) | 2.6 (0.9-4.4) | 0.0013 |
| Systolic BP ≥ 140.0 | 5010 / 1218 (24.3%) | 58.7 (56.9-60.5) | 63.5 (61.7-65.2) | 4.8 (3.1-6.4) | 0.0000 |
| Total bilirubin > 1.0 | 3212 / 376 (11.7%) | 64.3 (61.5-67.2) | 72.0 (69.3-74.7) | 7.7 (5.1-10.3) | 0.0000 |
| Total cholesterol ≥ 200.0 | 3402 / 1099 (32.3%) | 59.6 (57.6-61.6) | 61.2 (59.1-63.2) | 1.6 (-0.5-3.6) | 0.0672 |
| Total cholesterol ≥ 240.0 | 3402 / 377 (11.1%) | 58.3 (55.4-61.2) | 62.3 (59.3-65.4) | 4.0 (0.6-7.4) | 0.0107 |
| Triglycerides ≥ 150.0 | 3369 / 1511 (44.9%) | 57.8 (55.9-59.7) | 62.8 (61.0-64.7) | 5.0 (3.2-6.9) | 0.0000 |
| Triglycerides ≥ 200.0 | 3369 / 868 (25.8%) | 55.9 (53.7-58.1) | 62.3 (60.2-64.4) | 6.4 (4.6-8.3) | 0.0000 |
| Triglycerides ≥ 500.0 | 3369 / 79 (2.3%) | 63.6 (57.6-69.6) | 67.3 (61.8-72.8) | 3.7 (-0.4-7.9) | 0.0380 |
| TSH < 0.5 | 1034 / 67 (6.5%) | 58.7 (51.6-65.8) | 58.1 (51.3-64.9) | -0.6 (-7.3-6.1) | 0.5672 |
| TSH > 4.0 | 1034 / 103 (10.0%) | 58.9 (53.0-64.8) | 62.2 (56.9-67.5) | 3.3 (-2.8-9.3) | 0.1441 |
| WBC < 4.0 | 4731 / 96 (2.0%) | 60.0 (54.6-65.3) | 72.6 (67.4-77.8) | 12.6 (6.9-18.3) | 0.0000 |
| WBC > 11.0 | 4731 / 314 (6.6%) | 58.5 (55.2-61.9) | 61.6 (58.2-65.0) | 3.1 (-0.1-6.2) | 0.0294 |

**B**

| Prediction | n / N (%) | Baseline AUC (CI) | DLS AUC (CI) | Improvement (CI) | p |
|---|---|---|---|---|---|
| ACR ≥ 30.0 | 1765 / 556 (31.5%) | 55.4 (52.6-58.2) | 60.7 (57.9-63.5) | 5.3 (1.9-8.6) | 0.0010 |
| ACR ≥ 300.0 | 1765 / 151 (8.6%) | 53.2 (48.4-58.0) | 66.4 (62.0-70.9) | 13.2 (7.0-19.4) | 0.0000 |
| ACR ≥ 1500.0 | 1765 / 37 (2.1%) | 49.1 (38.9-59.4) | 67.2 (59.3-75.0) | 18.0 (5.0-31.0) | 0.0033 |
| BMI ≥ 25.0 | 4783 / 4033 (84.3%) | 54.8 (52.4-57.1) | 72.0 (70.1-73.9) | 17.3 (14.5-20.0) | 0.0000 |
| BMI ≥ 30.0 | 4783 / 2399 (50.2%) | 57.6 (56.0-59.3) | 70.3 (68.9-71.8) | 12.7 (10.9-14.4) | 0.0000 |
| BMI ≥ 35.0 | 4783 / 1018 (21.3%) | 56.8 (54.9-58.7) | 71.1 (69.3-72.8) | 14.2 (12.2-16.2) | 0.0000 |
| BMI ≥ 40.0 | 4783 / 372 (7.8%) | 59.5 (56.8-62.2) | 74.5 (72.0-77.0) | 15.0 (12.2-17.9) | 0.0000 |
| Creatinine > 1.2 | 4898 / 1401 (28.6%) | 68.5 (66.9-70.1) | 66.9 (65.3-68.5) | -1.6 (-3.2--0.1) | 0.9833 |
| Diastolic BP ≥ 80.0 | 5141 / 1994 (38.8%) | 61.7 (60.1-63.3) | 65.8 (64.2-67.3) | 4.1 (2.7-5.4) | 0.0000 |
| Diastolic BP ≥ 90.0 | 5141 / 364 (7.1%) | 66.3 (63.4-69.2) | 68.3 (65.5-71.0) | 1.9 (-0.6-4.4) | 0.0659 |
| eGFR < 15.0 | 4898 / 38 (0.8%) | 61.3 (53.5-69.2) | 79.0 (71.8-86.2) | 17.7 (6.0-29.3) | 0.0015 |
| eGFR < 30.0 | 4898 / 88 (1.8%) | 67.4 (62.4-72.4) | 79.2 (74.8-83.6) | 11.8 (5.4-18.2) | 0.0002 |
| eGFR < 60.0 | 4898 / 946 (19.3%) | 74.0 (72.4-75.7) | 73.7 (72.0-75.4) | -0.4 (-1.9-1.1) | 0.6874 |
| eGFR < 90.0 | 4898 / 3605 (73.6%) | 72.3 (70.7-73.8) | 70.5 (68.9-72.2) | -1.7 (-2.8--0.6) | 0.9987 |
| HbA1c ≥ 6.5 | 3089 / 1519 (49.2%) | 47.4 (45.4-49.5) | 55.9 (53.9-57.9) | 8.5 (6.7-10.3) | 0.0000 |
| HbA1c ≥ 7.0 | 3089 / 1110 (35.9%) | 47.9 (45.8-50.0) | 59.6 (57.5-61.6) | 11.7 (9.5-13.8) | 0.0000 |
| HbA1c ≥ 8.0 | 3089 / 604 (19.6%) | 51.2 (48.6-53.7) | 60.5 (58.1-63.0) | 9.4 (6.8-11.9) | 0.0000 |
| HbA1c ≥ 9.0 | 3089 / 347 (11.2%) | 60.1 (57.1-63.0) | 65.7 (62.7-68.7) | 5.6 (2.7-8.5) | 0.0001 |
| HCT < 39.0 | 4601 / 759 (16.5%) | 68.9 (66.9-71.0) | 72.6 (70.6-74.5) | 3.6 (1.2-6.0) | 0.0014 |
| HDL ≥ 45.0 | 4442 / 2183 (49.1%) | 62.7 (61.0-64.3) | 63.6 (61.9-65.2) | 0.9 (-0.4-2.2) | 0.0925 |
| HDL ≥ 60.0 | 4442 / 725 (16.3%) | 62.5 (60.2-64.8) | 64.9 (62.8-67.1) | 2.4 (0.5-4.4) | 0.0080 |
| Hgb < 11.0 | 4601 / 180 (3.9%) | 66.5 (62.6-70.3) | 73.8 (70.4-77.2) | 7.3 (2.1-12.6) | 0.0031 |
| Hgb < 12.5 | 4601 / 689 (15.0%) | 70.2 (68.0-72.3) | 73.8 (71.9-75.7) | 3.7 (1.2-6.2) | 0.0021 |
| LDL ≥ 100.0 | 4347 / 2059 (47.4%) | 63.7 (62.0-65.3) | 62.4 (60.8-64.1) | -1.3 (-2.8-0.2) | 0.9511 |
| LDL ≥ 130.0 | 4347 / 914 (21.0%) | 60.6 (58.6-62.6) | 59.1 (57.1-61.1) | -1.5 (-3.6-0.6) | 0.9168 |
| LDL ≥ 160.0 | 4347 / 288 (6.6%) | 59.4 (56.2-62.7) | 57.3 (53.8-60.7) | -2.2 (-6.0-1.7) | 0.8674 |
| LDL ≥ 190.0 | 4347 / 77 (1.8%) | 61.8 (55.7-68.0) | 59.8 (53.6-66.0) | -2.0 (-8.4-4.3) | 0.7343 |
| Mean arterial pressure ≥ 110.0 | 5141 / 308 (6.0%) | 60.6 (57.2-64.1) | 64.8 (61.7-67.9) | 4.2 (0.9-7.5) | 0.0067 |
| Mean arterial pressure ≥ 80.0 | 5141 / 4917 (95.6%) | 54.4 (50.4-58.4) | 56.3 (52.5-60.1) | 1.9 (-3.4-7.2) | 0.2421 |
| Mean arterial pressure ≥ 90.0 | 5141 / 3756 (73.1%) | 58.4 (56.7-60.2) | 60.4 (58.7-62.2) | 2.0 (0.2-3.8) | 0.0135 |
| Platelet < 100.0 | 4597 / 42 (0.9%) | 63.2 (55.4-70.9) | 59.0 (49.9-68.1) | -4.2 (-12.8-4.4) | 0.8292 |
| Platelet < 150.0 | 4597 / 314 (6.8%) | 66.2 (63.3-69.0) | 58.9 (55.7-62.0) | -7.3 (-10.4--4.2) | 1.0000 |
| Pulse pressure ≥ 40.0 | 5141 / 4523 (88.0%) | 63.8 (61.5-66.2) | 64.5 (62.1-66.9) | 0.6 (-1.4-2.7) | 0.2689 |
| Pulse pressure ≥ 55.0 | 5141 / 2007 (39.0%) | 68.9 (67.4-70.3) | 67.7 (66.2-69.2) | -1.2 (-2.3--0.1) | 0.9834 |
| Pulse pressure ≥ 65.0 | 5141 / 851 (16.6%) | 71.5 (69.8-73.2) | 70.1 (68.4-71.9) | -1.4 (-2.7--0.0) | 0.9756 |
| Systolic BP ≥ 120.0 | 5141 / 3934 (76.5%) | 58.8 (56.9-60.7) | 59.9 (58.1-61.8) | 1.1 (-1.1-3.3) | 0.1565 |
| Systolic BP ≥ 140.0 | 5141 / 1305 (25.4%) | 60.4 (58.7-62.1) | 60.2 (58.5-61.9) | -0.2 (-2.1-1.8) | 0.5704 |
| Total cholesterol ≥ 200.0 | 4448 / 1132 (25.4%) | 60.6 (58.7-62.5) | 59.0 (57.1-60.8) | -1.7 (-3.9-0.5) | 0.9302 |
| Total cholesterol ≥ 240.0 | 4448 / 322 (7.2%) | 59.5 (56.3-62.8) | 60.5 (57.3-63.7) | 1.0 (-2.3-4.2) | 0.2771 |
| Triglycerides ≥ 150.0 | 4430 / 1282 (28.9%) | 62.5 (60.6-64.3) | 64.5 (62.7-66.3) | 2.0 (0.8-3.3) | 0.0006 |
| Triglycerides ≥ 200.0 | 4430 / 710 (16.0%) | 64.1 (61.9-66.3) | 65.3 (63.1-67.5) | 1.3 (-0.2-2.8) | 0.0487 |
| Triglycerides ≥ 500.0 | 4430 / 55 (1.2%) | 66.5 (59.1-74.0) | 64.5 (57.3-71.7) | -2.0 (-7.6-3.5) | 0.7638 |
| WBC < 4.0 | 4600 / 317 (6.9%) | 68.2 (65.6-70.8) | 64.5 (61.5-67.4) | -3.7 (-6.8--0.6) | 0.9899 |

## C

| Prediction | n / N (%) | Baseline AUC (CI) | DLS AUC (CI) | Improvement (CI) | p |
|---|---|---|---|---|---|
| ACR ≥ 30.0 | 6717 / 2007 (29.9%) | 55.5 (54.0-57.0) | 62.1 (60.6-63.5) | 6.6 (4.8-8.3) | 0.0000 |
| ACR ≥ 300.0 | 6717 / 541 (8.1%) | 54.4 (51.9-56.9) | 66.1 (63.7-68.5) | 11.7 (8.6-14.9) | 0.0000 |
| ACR ≥ 1500.0 | 6717 / 132 (2.0%) | 54.0 (49.2-58.7) | 67.3 (62.0-72.6) | 13.3 (6.3-20.4) | 0.0001 |
| Creatinine > 1.2 | 9903 / 3209 (32.4%) | 63.7 (62.5-64.8) | 63.3 (62.2-64.5) | -0.3 (-1.5-0.8) | 0.7306 |
| eGFR < 15.0 | 9903 / 91 (0.9%) | 50.9 (45.0-56.7) | 80.0 (75.7-84.3) | 29.1 (21.7-36.6) | 0.0000 |
| eGFR < 30.0 | 9903 / 266 (2.7%) | 59.9 (56.7-63.1) | 76.9 (74.2-79.6) | 17.0 (13.2-20.9) | 0.0000 |
| eGFR < 60.0 | 9903 / 2311 (23.3%) | 70.5 (69.3-71.7) | 71.6 (70.4-72.7) | 1.1 (-0.0-2.1) | 0.0288 |
| eGFR < 90.0 | 9903 / 7139 (72.1%) | 70.6 (69.5-71.7) | 70.7 (69.6-71.8) | 0.1 (-0.9-1.1) | 0.4050 |
| HbA1c ≥ 6.5 | 9255 / 6913 (74.7%) | 56.5 (55.1-57.8) | 60.1 (58.8-61.5) | 3.7 (2.0-5.3) | 0.0000 |
| HbA1c ≥ 7.0 | 9255 / 5282 (57.1%) | 57.7 (56.5-58.9) | 61.7 (60.5-62.8) | 4.0 (2.5-5.5) | 0.0000 |
| HbA1c ≥ 8.0 | 9255 / 3045 (32.9%) | 61.9 (60.7-63.1) | 65.3 (64.1-66.5) | 3.4 (2.1-4.8) | 0.0000 |
| HbA1c ≥ 9.0 | 9255 / 1843 (19.9%) | 64.9 (63.6-66.3) | 68.0 (66.6-69.3) | 3.0 (1.7-4.4) | 0.0000 |
| HDL ≥ 45.0 | 9575 / 3219 (33.6%) | 53.4 (52.2-54.7) | 63.1 (61.9-64.3) | 9.7 (8.1-11.3) | 0.0000 |
| HDL ≥ 60.0 | 9575 / 695 (7.3%) | 53.7 (51.3-56.1) | 68.8 (66.8-70.8) | 15.1 (12.3-17.8) | 0.0000 |
| LDL ≥ 100.0 | 9282 / 3466 (37.3%) | 61.2 (60.0-62.4) | 62.2 (61.0-63.4) | 1.0 (0.0-2.1) | 0.0247 |
| LDL ≥ 130.0 | 9282 / 1503 (16.2%) | 60.9 (59.3-62.5) | 61.8 (60.3-63.3) | 0.9 (-0.6-2.3) | 0.1213 |
| LDL ≥ 160.0 | 9282 / 495 (5.3%) | 60.7 (58.1-63.3) | 58.7 (56.2-61.2) | -2.0 (-4.5-0.4) | 0.9511 |
| LDL ≥ 190.0 | 9282 / 155 (1.7%) | 63.9 (59.4-68.5) | 61.1 (56.5-65.6) | -2.9 (-7.0-1.2) | 0.9141 |
| Total cholesterol ≥ 200.0 | 9588 / 1986 (20.7%) | 61.0 (59.6-62.4) | 59.9 (58.5-61.3) | -1.1 (-2.5-0.3) | 0.9391 |
| Total cholesterol ≥ 240.0 | 9588 / 610 (6.4%) | 62.5 (60.2-64.8) | 61.5 (59.3-63.8) | -1.0 (-3.0-1.0) | 0.8281 |
| Triglycerides ≥ 150.0 | 9516 / 3971 (41.7%) | 52.5 (51.3-53.6) | 65.8 (64.7-66.9) | 13.4 (11.9-14.8) | 0.0000 |
| Triglycerides ≥ 200.0 | 9516 / 2373 (24.9%) | 55.0 (53.7-56.3) | 66.1 (64.9-67.4) | 11.2 (9.6-12.7) | 0.0000 |
| Triglycerides ≥ 500.0 | 9516 / 255 (2.7%) | 66.0 (62.7-69.2) | 68.3 (65.1-71.4) | 2.3 (-0.8-5.4) | 0.0742 |

**D**

| Variable (A-H) | Units | Variable (L-W) | Units |
|---|---|---|---|
| ACR | mg/g | LDL | mg/dL |
| Albumin | g/dL | Mean arterial pressure | mmHg |
| ALT | U/L | non-HDL | mg/dL |
| AST | U/L | Platelet | $10^3/\mu L$ |
| BMI | $kg/m^2$ | Potassium | mEq/L |
| BUN | mg/dL | Pulse pressure | mmHg |
| Calcium | mg/dL | RDW | % |
| Creatinine | mg/dL | Sodium | mEq/L |
| Diastolic BP | mmHg | Systolic BP | mmHg |
| eGFR | $mL/min/1.73\ m^2$ | Total bilirubin | mg/dL |
| HgbA1c | % | Total cholesterol | mg/dL |
| HCT | % | Triglycerides | mg/dL |
| HDL | mg/dL | TSH | mU/L |
| Hgb | g/dL | WBC | $10^3/\mu L$ |

**Supplementary Table 6. Adjusted analysis**

Reference category for race is White. DLS score was normalized to have unit variance and zero mean (i.e. z-score normalization). YrsDM indicates "years with diabetes". For numerical stability of the estimate, we pooled race/ethnicity groups comprising < 2% of the population into "Race=Other". Bold indicates p<0.05. †Indicates that the target was prespecified as secondary analysis; all others were prespecified as primary analysis.

**A**

| Prediction | Variable | Odds ratio (CI) | p |
|---|---|---|---|
| eGFR < 60.0 | **Age** | **0.979 (0.969-0.989)** | **0.0001** |
| | **Sex = Male** | **0.576 (0.471-0.705)** | **0.0000** |
| | Race=Black | 0.922 (0.560-1.517) | 0.7482 |
| | Race=Asian / Pacific islander | 1.058 (0.623-1.799) | 0.8345 |
| | **Race=Hispanic** | **0.496 (0.323-0.760)** | **0.0013** |
| | Race=Other | 0.937 (0.536-1.637) | 0.8184 |
| | **YrsDM** | **1.034 (1.018-1.050)** | **0.0000** |
| | **DLS** | **2.318 (2.113-2.542)** | **0.0000** |
| Hgb < 11.0 | **Age** | **0.973 (0.964-0.983)** | **0.0000** |
| | **Sex=Male** | **0.688 (0.552-0.856)** | **0.0008** |
| | Race=Black | 1.051 (0.561-1.968) | 0.8770 |
| | Race=Asian / Pacific islander | 0.938 (0.466-1.891) | 0.8588 |
| | **Race=Hispanic** | **0.505 (0.290-0.879)** | **0.0157** |
| | Race=Other | 0.849 (0.412-1.750) | 0.6575 |
| | **YrsDM** | **1.027 (1.010-1.045)** | **0.0019** |
| | **DLS** | **2.328 (2.156-2.513)** | **0.0000** |

**B**

| Prediction | Variable | Odds ratio (CI) | p |
|---|---|---|---|
| eGFR < 30.0† | **Age** | **1.060 (1.036-1.085)** | **0.0000** |
| | **Sex=Male** | **0.489 (0.243-0.982)** | **0.0444** |
| | **Race=Black** | **1.911 (1.187-3.077)** | **0.0077** |
| | Race=Other | 0.477 (0.020-11.217) | 0.6461 |
| | **DLS** | **1.733 (1.538-1.953)** | **0.0000** |
| Hgb < 11.0 | Age | 1.002 (0.988-1.016) | 0.7723 |
| | **Sex=Male** | **0.269 (0.188-0.384)** | **0.0000** |
| | **Race=Black** | **1.584 (1.111-2.259)** | **0.0110** |
| | Race=Other | 1.176 (0.280-4.939) | 0.8252 |
| | **DLS** | **1.904 (1.698-2.134)** | **0.0000** |

**C**

| Prediction | Variable | Odds ratio (CI) | p |
|---|---|---|---|
| eGFR < 30.0† | Age | 0.984 (0.962-1.005) | 0.1402 |
| | **Sex=Male** | **0.324 (0.166-0.634)** | **0.0010** |
| | Race=Black | 0.874 (0.564-1.355) | 0.5480 |
| | Race=Other | 0.989 (0.230-4.261) | 0.9887 |
| | **DLS** | **1.827 (1.639-2.037)** | **0.0000** |

**Supplementary Table 7. Hyperparameters**

(A) common across models and (B) specific to each selected model (i.e. chosen via hyperparameter search).

**A**

| Hyperparameter | Value |
|---|---|
| Batch size per TPU core | 8 |
| Number of TPU cores | 8 |
| Effective batch size | 64 |
| Optimizer | Momentum |
| Warm up steps | 2000 |
| Moving average decay rate | 0.9999 |
| Max training steps | 50000 |
| Input image size [pixels] | 587 × 587 |
| Random flip | Horizontal and vertical |
| Random brightness max delta | 0.1148 |
| Random hue max delta | 0.0251 |
| Random saturation range | 0.5597 - 1.2749 |
| Random contrast range | 0.9997 - 1.7705 |
| Random scale factors | 1.0X (60%), 1.3X (20%), 1.5X (20%) |
| Weight decay rate (L2 regularization) | 0.0 |
| Model architecture | ResNet BiT-M-R101x3 |

**B**

| Hyperparameter | Model 1 | Model 2 | Model 3 | Model 4 | Model 5 |
|---|---|---|---|---|---|
| Initial learning rate | 3.49e-03 | 5.67e-03 | 1.79e-02 | 7.38e-03 | 1.30e-02 |
| Pre warm-up learning rate | 3.49e-04 | 5.67e-04 | 1.79e-03 | 7.38e-04 | 1.30e-03 |
| Decay steps | 10000 | 10000 | 10000 | 20000 | 10000 |
| Decay rate | 0.306 | 0.615 | 0.153 | 0.595 | 0.962 |
| Training steps (early stopping) | 8000 | 8000 | 14000 | 10000 | 14000 |
| Dropout rate at final layer | 0.13 | 0.15 | 0.35 | 0.47 | 0.73 |
| Smoking aux loss | No | No | Yes | No | Yes |
| Eye disease aux loss | No | Yes | Yes | No | No |